**Author for correspondence:**
Pengzhi Lin
e-mail: cvelinpz@scu.edu.cn

# The theory of fifth-order Stokes waves in a linear shear current

Haiqi Fang[1], Philip L.-F. Liu[2,3,4], Lian Tang[1] and Pengzhi Lin[1]

[1]State Key Laboratory of Hydraulics and Mountain River Engineering, Sichuan University, Chengdu, 610065, China
[2]School of Civil and Environmental Engineering, Cornell University, Ithaca, NY, 14853, USA
[3]Institute of Hydrological and Oceanic Sciences, National Central University, Taoyuan City, 32001, Taiwan
[4]Department of Hydraulic and Ocean Engineering, National Cheng Kung University, Tainan City, 70101, Taiwan

In this study, a new set of fifth-order Stokes wave solutions, incorporating the effects of a linear shear current, is derived by utilizing the perturbation method originally proposed for pure waves that was recently published. The present solutions are checked against the existing experimental data, the third-order stream function solutions, as well as the numerical results. The comparisons demonstrate that the present solutions are more accurate in describing the velocity distributions during wave propagation, especially in strong following currents and positive vorticity conditions. Subsequently, the present solutions are used to investigate the fluid particle trajectories for different wave-current interaction conditions. The results indicate that the background vorticity can alter the patterns of fluid particle trajectories and the direction of Stokes drifts.

## 1. Introduction

In coastal areas, shear currents, accompanied by vorticity, are often generated by wind action and bottom friction. In deep water, wind stress generates a thin wind-draft layer, causing a surface drift of the water. This leads to a strong vertical shearing action within the upper layer of the water body. Therefore, the rotational characteristics of the flow motion play an essential role in affecting the wave-current interaction process. A comprehensive overview of wave-current interaction can be found in [1,2].

The previous research has highlighted the significant impact of a vertically varying current that can substantially alter the kinematic and dynamic properties of surface water waves. First and foremost, based on the stream function theory, Dalrymple [3] introduced a Fourier expansion technique to investigate the nonlinear propagation of steady waves under a linear shear current condition, providing a viable numerical methodology for studying wave-current interaction. Simmen & Saffman [4] adopted a boundary-integral method to examine steep waves in deep water, which was further extended to finite-depth water by Silva & Peregrine [5]. Their study revealed that the ratio between kinetic energy and potential energy in the presence of vorticity shows notable difference from the that for the pure wave theory. Additionally, Moreira & Chacaltana [6] identified that the breaking limit is substantially influenced by vorticity. Moreover, Francius & Kharif [7] demonstrated that linear shear currents can alter the Benjamin-Feir modulational instabilities of waves. By extending the Fourier approximation method of Rienecker & Fenton [8], Francius & Kharif [7] included the effects of linear shear currents for highly nonlinear waves. Recently, Murashige & Choi [9] further investigated the effects of linear shear currents on the instability of nonlinear waves using an unsteady conformal mapping.

Analytical Stokes wave solutions provide information on wave properties and velocity fields for ocean and coastal engineering applications, enabling extensive studies of wave dynamics. Since the pioneering work of Stokes [10], subsequent researchers have presented numerous perturbation solutions [11–15], which generates extensive studies on the nonlinear characteristics of waves [16–19]. Based on the stream function expansions, Fenton [12] introduced a set of fifth-order solutions, which has been widely utilized in coastal and ocean engineering community as a reliable solution. However, Zhao & Liu [15] recently conducted a detailed investigation into the fifth-order Stokes wave solutions. They pointed that the assumption made by Fenton [12], requiring the wave height of the solutions be twice pf the wave amplitude of the first order and first harmonic solution $(H = 2A)$, was non-physical. Zhao & Liu [15] employed an alternative method by introducing a perturbation expansion to the wave speed and derived a new set of fifth-order Stokes wave solutions for pure waves.

On the other hand, to integrate the impact of linear shear current into Stokes wave solutions, Tsao [20] proposed a set of third-order Stokes wave solutions, which was subsequently extended to deep water by Brevik [21]. Following the work of Fenton [12], Kishida & Sobey [22] incorporated current effects into the third-order Stokes solutions, which were further extended to include surface tension to study the capillary gravity waves [23,24]. Nevertheless, the derivations of these solutions are all based on Fenton's [12] non-physical assumption, $(H = 2A)$, as mentioned above. Thus, the existing solutions cannot remain valid when the current and wave nonlinearity are considerably strong. Moreover, to the best of the authors' knowledge, all the existing Stokes wave solutions, considering the effect of linear shear current, are only up to the third-order. However, experimental evidence [25] showed that higher order solutions are important in predicting wave profiles under a strongly sheared adverse current condition. Therefore, to better understand wave properties in a linear shear current and better serve engineering applications, a more physical and accurate fifth-order solution is still in wanting.

In this paper, based on the perturbation method proposed by Zhao & Liu [15] for deriving the fifth-order Stokes wave solutions, a set of fifth-order Stokes wave solutions for waves in a linear shear current is proposed. In section 2, the governing equations, boundary conditions and the perturbation method solution process are briefly summarized. In section 3, the accuracy of the proposed solutions is validated against existing experimental data and numerical solutions of Francius & Kharif [7]. In section 4, the capabilities of the present solutions in predicting waves with intense nonlinearity, strong surface current, and large vorticity are examined by comparing them with the numerical results [7] and the third-order solutions of Kishida & Sobey [22], followed by the investigation of the properties of fluid particle trajectories under different current conditions. In section 5, the main conclusions and results are outlined. In the electronic supplementary material, the detailed derivation of the perturbation expansion is presented.

## 2. Derivation of fifth-order Stokes wave solution in a linear shear current

In this section, we present a derivation of fifth-order Stokes wave solutions in a linear shear current, based on the framework of Zhao & Liu [15]. The governing equations and the boundary conditions are first presented, being followed by the detailed perturbation method solution process.

### (a) Governing equations

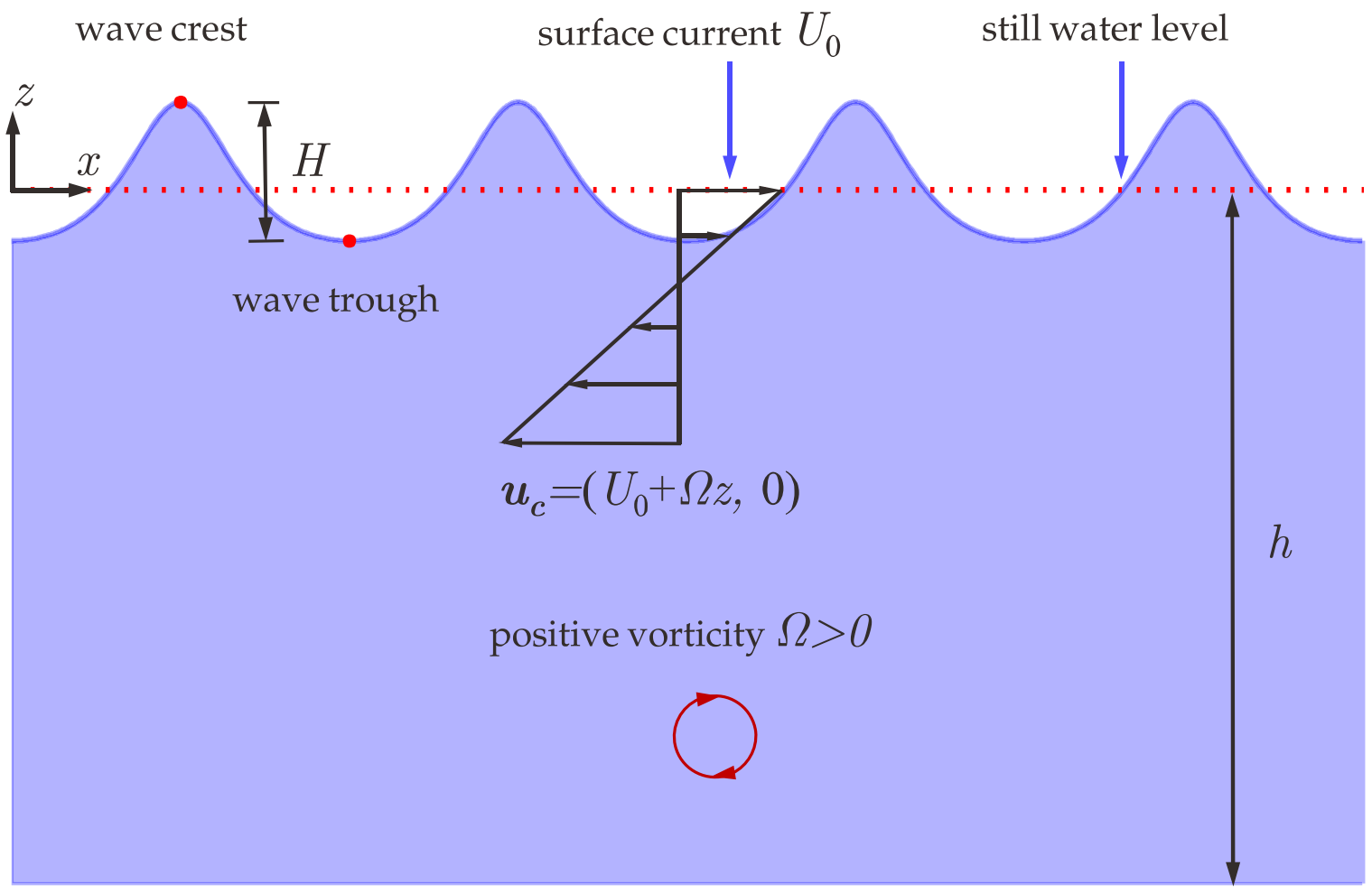

**Figure 1.** A definition sketch for a periodic Stokes wave propagation with a vertical linear shear current.

Figure 1 depicts a periodic Stokes wave propagating along the positive horizontal $x$-axis direction (from left to right) with a vertical, linearly distributed sheared current on a constant water depth $h$. The vertical $z$-axis is positive upward with $z = 0$ being at the still water depth. $H$ denotes the wave height, the vertical distance between wave crest and wave trough. $\boldsymbol{u_c}$ represents the current vector,

$$\boldsymbol{u_c} = (U_0 + \Omega z, 0) \tag{2.1}$$

where $U_0$ and $\Omega$ represent the surface current (at $z = 0$) and constant vorticity, respectively. The clockwise arrow of the red circle indicates the positive direction of shear, with $\Omega > 0$, and the reverse direction represents a negative shear ($\Omega < 0$). $\bar{U}$ is the depth-averaged current, which is defined as:

$$\bar{U} = U_0 - \frac{1}{2}\Omega h \tag{2.2}$$

The total velocity field, $\boldsymbol{u}$, can be decomposed into two parts, namely, the wave-induced velocity and current, $\boldsymbol{u} = \boldsymbol{u_w} + \boldsymbol{u_c}$. Generally, the wave-induced velocity field with vorticity cannot be described by a potential theory. However, since the linear shear current is divergence free, the wave-induced velocity can be represented by a potential function $\phi(x,z,t)$, i.e., $\boldsymbol{u_w} = \nabla\phi = (u_w, w_w)$. The governing equation and boundary conditions for the velocity potential can be specified as (e.g., [4,26]),

$$\phi_{xx} + \phi_{zz} = 0. \quad -h \le z \le \eta \tag{2.3}$$

On the free surface, the dynamic free surface condition is

$$\phi_t + g\eta + (U_0 + \Omega\eta)\,\phi_x + \frac{1}{2}\left(\phi_x^2 + \phi_z^2\right) - \Omega\psi = 0, \quad z = \eta \tag{2.4}$$

and the kinematic boundary condition is

$$\phi_z = \eta_t + \eta_x\left(U_0 + \Omega\eta + \phi_x\right), \quad z = \eta \tag{2.5}$$

in which $g$ represents the gravitational acceleration and $\eta(x,t)$ is the free surface elevation. On the free surface boundary condition, (2.4), $\psi$ is the stream function, which can be related to the potential function as:

$$\psi = \int_{-h}^{z} \phi_x(x,\xi,t)d\xi. \tag{2.6}$$

At the sea bottom, the no-flux boundary condition is

$$\phi_z = 0. \quad z = -h \tag{2.7}$$

Equations (2.3)-(2.7) constitute the governing equations and the corresponding boundary conditions that govern the motion of waves in linear shear currents, more details of the derivation can be found in [26].

## (b) Perturbation solutions for fifth-order Stokes wave

In this section, the derivation of the fifth-order Stokes wave considering a linear shear current by the perturbation method is presented. The dynamic and kinematic boundary conditions, equations (2.4) and (2.5), on the free surface, are first combined as

$$\left(\frac{\partial}{\partial t} + (U_0 + \Omega\eta + \phi_x)\frac{\partial}{\partial x} + \phi_z\frac{\partial}{\partial z}\right)\left[\phi_t + \frac{1}{2}\left(\phi_x^2 + \phi_z^2\right) + (U_0 + \Omega\eta)\,\phi_x - \Omega\psi\right] = -g\phi_z \tag{2.8}$$

which is applied on $z = \eta$. Appling the Taylor expansion of the above boundary condition at $z = 0$ yields:

$$\begin{aligned}\sum_{n=0}^{\infty}\frac{\eta^n}{n!}\frac{\partial^n}{\partial z^n}\Bigg\{g\phi_z + \left(\frac{\partial}{\partial t} + (U_0 + \Omega\eta + \phi_x)\frac{\partial}{\partial x} + \phi_z\frac{\partial}{\partial z}\right) \times \\ \left[\phi_t + \frac{1}{2}\left(\phi_x^2 + \phi_z^2\right) + (U_0 + \Omega\eta)\,\phi_x - \Omega\psi\right]\Bigg\} = 0. \quad z = 0\end{aligned} \tag{2.9}$$

Similarly, the free surface elevation can be obtained by employing the Taylor expansion to the dynamic boundary condition, (2.4):

$$\eta = -\frac{1}{g}\sum_{n=0}^{\infty}\frac{\eta^n}{n!}\frac{\partial^n}{\partial z^n}\left\{\phi_t + (U_0 + \Omega\eta)\,\phi_x + \frac{1}{2}\left(\phi_x^2 + \phi_z^2\right) - \Omega\psi\right\}. \quad z = 0 \tag{2.10}$$

Assuming the solutions for equations (2.3)-(2.7) to be periodic in both time and space, which can be expressed in terms of different harmonics, $\sin j\theta$ and $\cos j\theta (j \geq 0)$, where $\theta$ is the phase function:

$$\theta = kx - \omega t, \tag{2.11}$$

where $k$ and $\omega$ are the wave number and the wave angular frequency, respectively. The wave speed $c$ is defined as:

$$c = \frac{\omega}{k}. \tag{2.12}$$

According to Zhao & Liu [15], to ensure the solvability of the perturbation approach, a perturbation expansion of the wave angular frequency is required:

$$\omega = \left(1 + \sum_{i=1}^{\infty} \varepsilon^{(2\times i)} \beta_{(2\times i)}\right) \omega_0, \tag{2.13}$$

where $\varepsilon$ is an ordering parameter which will be set to unity in the final solutions, $\omega_0 = \sqrt{gk\sigma}$ and $\sigma = \tanh kh$. Subsequently, we introduce a stretched time variable $\tau$ to replace the original time variable $t$ :

$$\tau = \beta t, \tag{2.14}$$

and

$$\beta = 1 + \sum_{i=1}^{\infty} \varepsilon^{(2\times i)} \beta_{(2\times i)}. \tag{2.15}$$

where $\beta_2 (i = 1)$ and $\beta_4 (i = 2)$ are quadratic and quartic monomial of $kA$, respectively. By applying the substitution shown in (2.14), equations (2.11) and (2.12) can be rewritten as:

$$\theta = kx - \omega_0 \tau. \tag{2.16}$$

and

$$c = \frac{\beta \omega_0}{k}. \tag{2.17}$$

Then, the differential operator for time variable can be written as:

$$\frac{\partial}{\partial t} = \beta \frac{\partial}{\partial \tau}. \tag{2.18}$$

Thus, the combined and dynamic boundary condition, equations (2.9) and (2.10), can be expressed in the Taylor series form in terms of the new time variable $\tau$ :

$$\begin{aligned} \sum_{n=0}^{\infty} \frac{\eta^n}{n!} \frac{\partial^n}{\partial z^n} \Bigg\{ \beta^2 \phi_{\tau\tau} + g\phi_z + \left( \beta \frac{\partial}{\partial \tau} + (U_0 + \Omega\eta + \phi_x) \frac{\partial}{\partial x} + \phi_z \frac{\partial}{\partial z} \right) \times \\ \left[ \beta\phi_\tau + \frac{1}{2}\left(\phi_x^2 + \phi_z^2\right) + (U_0 + \Omega\eta)\, \phi_x - \Omega\psi \right] - \beta^2 \phi_{\tau\tau} \Bigg\} = 0, \quad z = 0 \end{aligned} \tag{2.19}$$

and

$$\eta = -\frac{1}{g} \sum_{n=0}^{\infty} \frac{\eta^n}{n!} \frac{\partial^n}{\partial z^n} \left\{ \beta\phi_\tau + (U_0 + \Omega\eta)\, \phi_x + \frac{1}{2}\left(\phi_x^2 + \phi_z^2\right) - \Omega\psi \right\}, \quad z = 0 \tag{2.20}$$

We seek for the following perturbation series forms of Stokes wave solutions, written as:

$$\phi = \sum_{i=1}^{\infty} \phi_i \varepsilon^i, \tag{2.21}$$

$$\psi = \sum_{i=1}^{\infty} \psi_i \varepsilon^i, \tag{2.22}$$

$$\eta = \sum_{i=1}^{\infty} \eta_i \varepsilon^i. \tag{2.23}$$

Substituting the series form solutions equations (2.21) - (2.23) into the governing equation (2.3) and the boundary conditions (2.7), (2.19) and (2.20), and then grouping the same terms with $\varepsilon^i$, the boundary value problems can be constructed and solved for $\phi^i, \eta^i$ and $\beta_{(2\times i)}$ (the derivations of each order problem and solution are based on Mathematica®); only the solutions up to fifth-order are discussed herein.

Following Zhao & Liu [15], the solutions for velocity potential, stream function as well as free surface profile can be formulated in the following form:

$$\begin{aligned}\phi(\theta,z) = &(A_{11}+A_{31}+A_{51})\cosh k(z+h)\sin\theta + \\ &(A_{22}+A_{42})\cosh 2k(z+h)\sin 2\theta + \\ &(A_{33}+A_{53})\cosh 3k(z+h)\sin 3\theta + \\ &A_{44}\cosh 4k(z+h)\sin 4\theta + A_{55}\cosh 5k(z+h)\sin 5\theta,\end{aligned} \tag{2.24}$$

$$\begin{aligned}\psi(\theta,z) = &(A_{11}+A_{31}+A_{51})\sinh k(z+h)\sin\theta + \\ &(A_{22}+A_{42})\sinh 2k(z+h)\sin 2\theta + \\ &(A_{33}+A_{53})\sinh 3k(z+h)\sin 3\theta + \\ &A_{44}\sinh 4k(z+h)\sin 4\theta + A_{55}\sinh 5k(z+h)\sin 5\theta,\end{aligned} \tag{2.25}$$

and

$$\begin{aligned}\eta(\theta) = &(B_{11}+B_{31}+B_{51})\cos\theta + (B_{22}+B_{42})\cos 2\theta + \\ &(B_{33}+B_{53})\cos 3\theta + B_{44}\cos 4\theta + B_{55}\cos 5\theta.\end{aligned} \tag{2.26}$$

And the dispersion equation can be expressed as:

$$\left(\omega\beta^{-1} - kU_0\right)\left(\omega\beta^{-1} - kU_0 + \Omega\sigma\right) = gk\sigma, \tag{2.27}$$

which incorporates the effects of the surface current $U_0$ and vorticity $\Omega$, and $\sigma$ is defined as $\tanh kh$. One should note that the dispersion equation is valid when the following conditions are satisfied: $\omega\beta^{-1} - kU_0 \geq 0$ and $\omega\beta^{-1} - kU_0 + \Omega\sigma \geq 0$.

The wave height $H$ is specified as

$$H = \eta(0) - \eta(\pi). \tag{2.28}$$

Given the prescribed values of $H, \omega, U_0$ and $\Omega$, equations (2.27) and (2.28) can be solved simultaneously for $k$ and $A$. On the other hand, if $H, k, U_0$ and $\Omega$ are specified, the solutions for $\omega$ and $A$ can be obtained. The details of the derivation for the fifth-order solutions can be found in the electronic supplementary material.

To evaluate the validity of the Stokes solutions under wave-current conditions, a new Ursell number is introduced, following Hedges [27] for pure waves,

$$\text{Ursell}^* = \left(\frac{2\pi}{kh}\right)^2 \frac{H}{h}\left(1+\gamma+\frac{1}{3}\gamma^2\right), \tag{2.29}$$

which is the ratio of second-order term of the potential function to the first-order term, and

$$\gamma = \frac{\Omega\tanh kh}{\omega - kU_0} = \frac{\Omega\tanh kh}{\omega - k\left(\bar{U} + \frac{1}{2}\Omega h\right)}, \tag{2.30}$$

which quantifies the effects of averaged current and vorticity. Under the no current condition, $\gamma = 0$, and Ursell* reduces to the Ursell number in Hedges [27], i.e., $4\pi^2 H/\left(k^2h^3\right)$. In this paper, if not specified, the Ursell number refers to Ursell*, equation (2.29).

To evaluate the effects of the averaged current and the vorticity, two dimensionless quantities are introduced to represent the ratio of the averaged current to wave speed and the ratio of vorticity to wave frequency, respectively,

$$U' = \frac{\bar{U}}{c}, \tag{2.31}$$

and

$$\Omega' = \frac{\Omega}{\omega}. \tag{2.32}$$

# 3. Solution validation

In this section, the proposed fifth-order Stokes wave solutions for different current conditions, namely, no current, uniform current and linear shear current, are presented and validated against the numerical solutions of Francius & Kharif [7], which employed the spectral method. In this paper, if not specified, the numerical solutions of Francius & Kharif [7] are referred to as the numerical results for brevity.

## (a) Fifth-order solutions for pure Stokes wave

For Stokes waves without current, the present solutions reduce to those of Zhao & Liu [15] for pure waves, which exhibit noticeable differences to Fenton's [12] solutions in predicting the wave speed, as shown in figure 2.

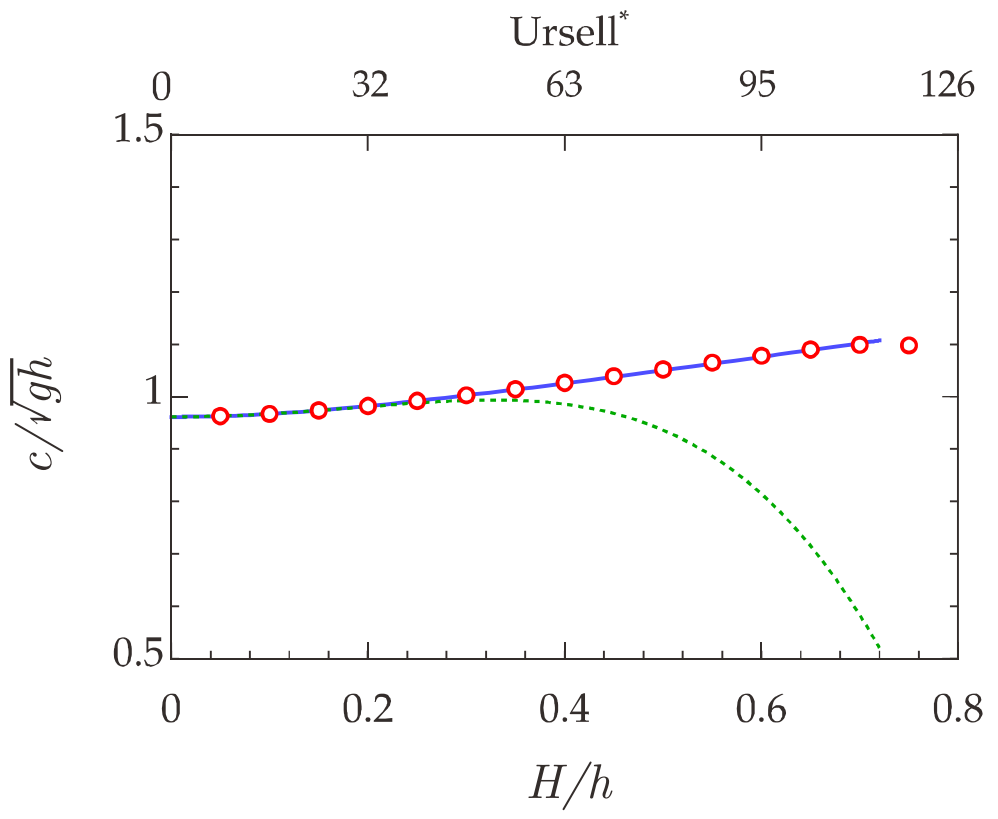

**Figure 2.** For the case of pure wave with $kh = 0.5$, the relationship between the dimensionless phase velocity, $c/\sqrt{gh}$, and the dimensionless nonlinearity $H/h$ is compared among the fifth-order solutions of Zhao & Liu [15] (blue solid line), the fifth-order solutions of Fenton [12] (green dotted line) and the numerical solutions of Francius & Kharif [7] (red circle).

Figure 2 illustrates the relationship between the dimensionless phase velocity $c/\sqrt{gh}$ and the dimensionless nonlinearity, $H/h$, when there is no current and $kh$ is fixed at $0.5$. When $H/h < 0.2$, the two analytical solutions of Zhao & Liu [15] and Fenton [12] are almost identical and are in satisfactory agreements with the numerical solutions. Increasing the wave nonlinearity, the phase speed predicted by Zhao & Liu [15] still shows good agreement with the numerical solutions, while the solutions of Fenton [12] diverge from the other two solutions and drop significantly with the further increase of dimensionless nonlinearity or Ursell number, demonstrating that the solutions of Zhao & Liu [15] are more accurate for strongly nonlinear waves, as the $H/h$ or Ursell value significantly increases.

## (b) Fifth-order solutions for Stokes wave with currents

In this section, the present fifth-order solutions with currents are validated against two sets of experiments reported in Thomas [28] and Swan [25], respectively, in which the horizontal current velocities were either uniform or linear shear.

### (i) Validation for uniform and weak linear shear current conditions

Thomas [28] conducted experiments to examine the impact of the vorticity on wavecurrent interaction. Four cases were performed with a fixed wave period, $T = 1.25$ s. Two of them considered uniform currents and the other two featured linear shear currents. The corresponding parameters are listed in table 1.

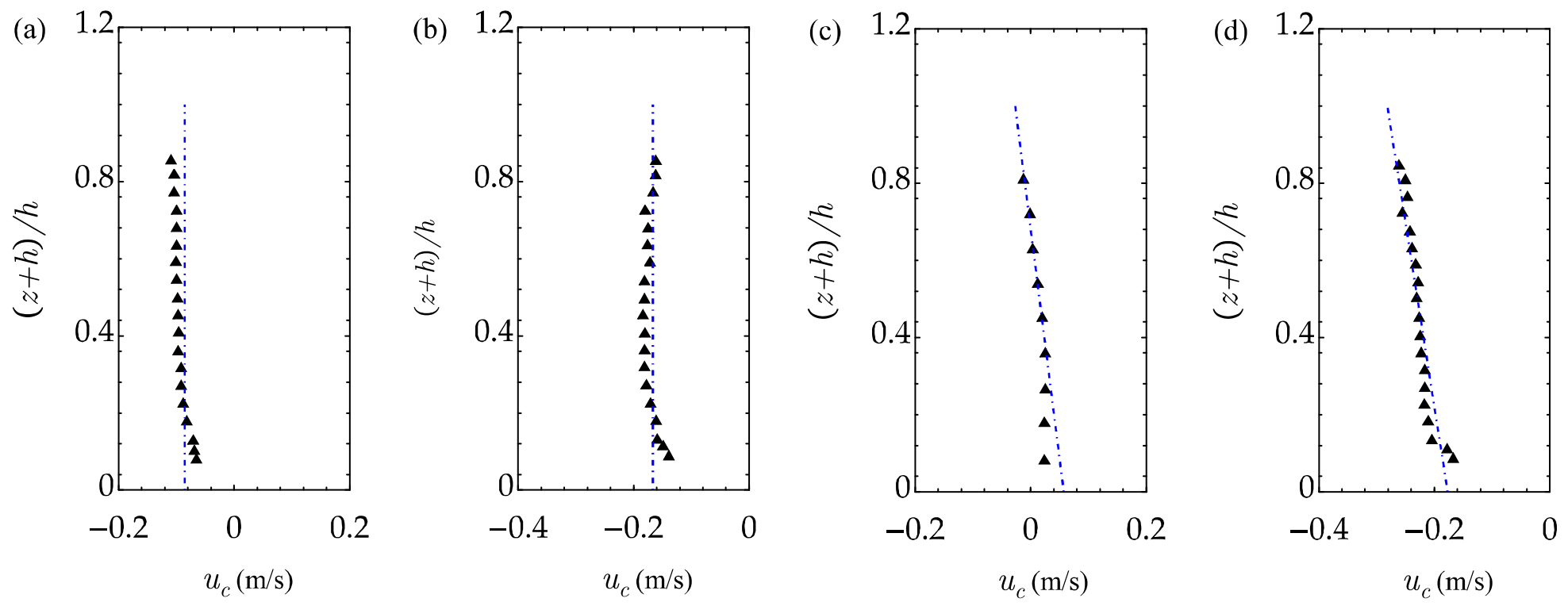

**Figure 3.** Comparisons between the current profile employed in the present solutions (blue dash-dotted line) and the measured current data by Thomas [28] (black triangle) for (a) case T1; (b) case T2; (c) case T3; (d) case T4.

**Table 1.** Parameters for wave and current in the experiment [28].

| case | $H$(mm) | $h$(m) | $T$(s) | $\bar{U}$(mm/s) | $\Omega$ (/s) | $H/h$ | $kh$ | $U'$ | $\Omega'$ | Ursell* |
|---|---|---|---|---|---|---|---|---|---|---|
| T1 | 137.78 | 0.548 | 1.25 | -85.7 | 0 | 0.25 | 1.6 | -0.05 | 0 | 3.85 |
| T2 | 144.84 | 0.545 | 1.25 | -166.5 | 0 | 0.27 | 1.7 | -0.10 | 0 | 3.54 |
| T3 | 129.02 | 0.551 | 1.25 | 15.3 | -0.15 | 0.23 | 1.5 | 0.01 | -0.03 | 3.95 |
| T4 | 151.46 | 0.549 | 1.25 | -229.3 | -0.19 | 0.28 | 1.9 | -0.16 | -0.04 | 2.92 |

Figure 3 depicts the current profiles by Thomas [28] as compared with the approximated current profiles for both uniform and weakly sheared current conditions, which are used in the present solutions. The corresponding horizontal velocity profiles in the water column under wave crest and wave trough are presented in figure 4. The horizontal velocity predicted by the present solutions shows consistency with the experimental results [28] for wave propagation under both uniform and weakly sheared current conditions. Slight differences are observed near the bottom, owing to the discrepancy between the experimentally measured currents and the approximated linear shear currents used in the analytical solutions. The bias of the horizontal velocity always shows the same direction with that of the current profile. Most importantly, the present solutions show good agreement with the numerical solutions, confirming the accuracy of the present and the numerical solutions for both uniform and linear shear currents.

### (ii) Validation for strong sheared current conditions

The predictive capability of the present solutions under strong vorticity conditions is checked by comparing them with the experimental data of Swan [25], in which the effects of strongly positive and negative vorticity on the wave velocity distribution were investigated experimentally. Two cases of the experiments are used in this analysis, as detailed in table 2 .

Figure 5 illustrates the measured current profiles as compared with the approximated linear shear current profiles for both opposing and following current conditions, which are used in the present solutions. The comparisons of the surface elevations are depicted in figure 6. Note that for case S1 (Ursell*=7.8) the surface elevation measured in experiments by Swan [25] does not show a perfect periodic variation in time (in figure 6 (a), the two wave crests are different and the small fluctuations appear at wave trough), which could be caused by measurement accuracy or the instability of the waves in the physical experiment, resulting in the discrepancy between the present solutions and the experimental data. Nevertheless, the surface elevation predicted by the present solutions match well with the numerical solutions, and the waveforms of both solutions present good stability and periodic variation. Moreover, the solutions of Touboul et al. [26], which was derived for linear wave propagation in linear shear currents, can also accurately describe waves under the conditions with small Ursell*. As for the case S2 with Ursell* $= 14.7$ (figure 6 (b)), the experimental data for the free surface elevations is not provided in [25]. However, the present solutions have satisfactory agreement with the numerical solutions, but have significant differences with the linear solution of Touboul et al. [26] due to the enhancement of the wave nonlinearity.

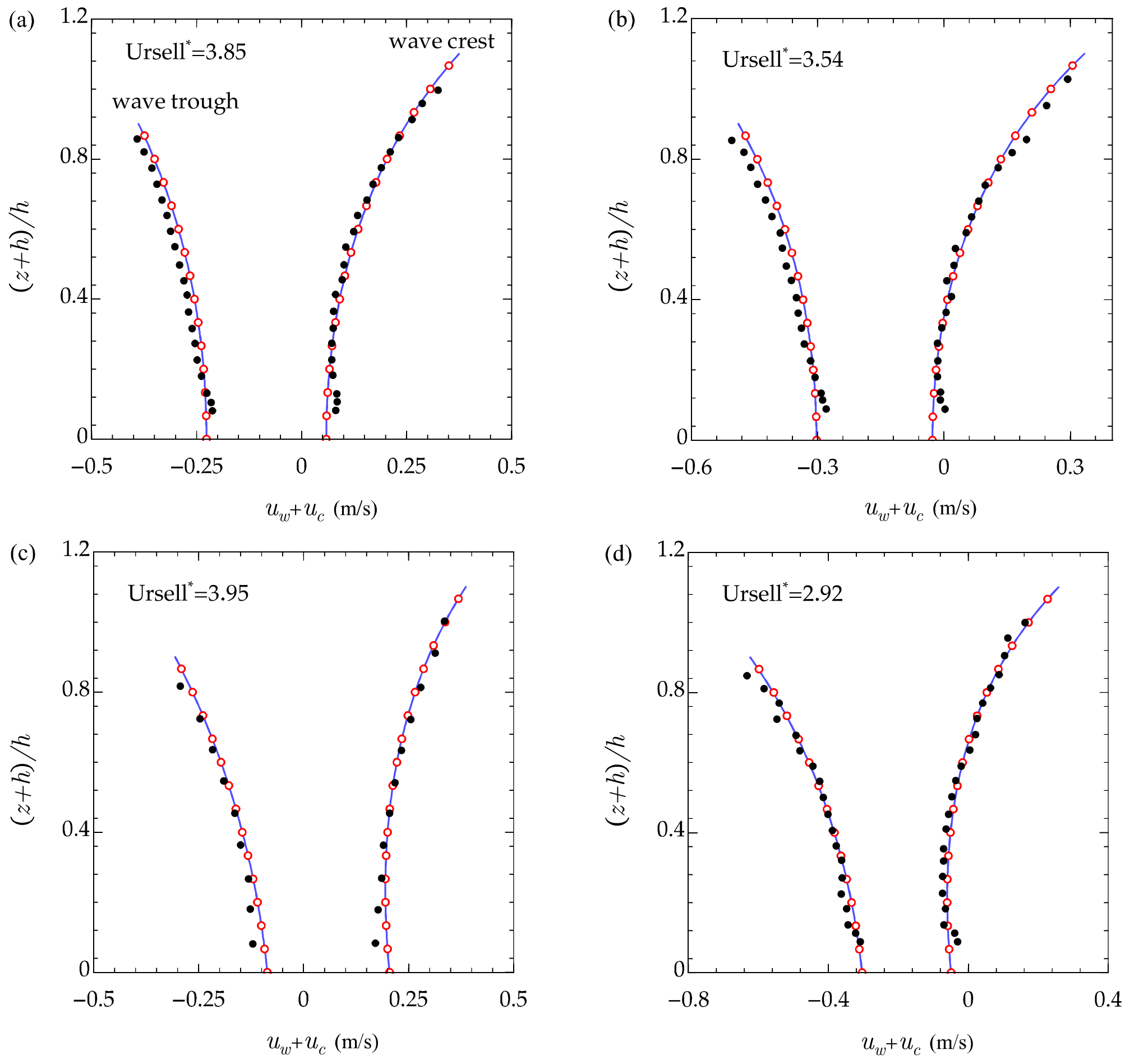

**Figure 4.** Comparisons of the horizontal velocities under the wave crest and wave trough among the present fifth-order solutions (blue solid line), the experimental data of Thomas [28] (black dot), and the numerical solutions of Francius & Kharif [7] (red circle) for (a) case T1; (b) case T2; (c) case T3; (d) case T4.

**Table 2.** Parameters for wave and current in the experiments [25].

| case | $H$(mm) | $h$(m) | $T$(s) | $\bar{U}$(m/s) | $\Omega$ (/s) | $H/h$ | $kh$ | $U'$ | $\Omega'$ | Ursell* |
|---|---|---|---|---|---|---|---|---|---|---|
| S1 | 123 | 0.35 | 1.420 | -0.208 | -1.67 | 0.35 | 1.2 | -0.16 | -0.38 | 7.8 |
| S2 | 63 | 0.35 | 1.418 | 0.123 | 1.70 | 0.1 | 0.8 | 0.06 | 0.38 | 14.7 |

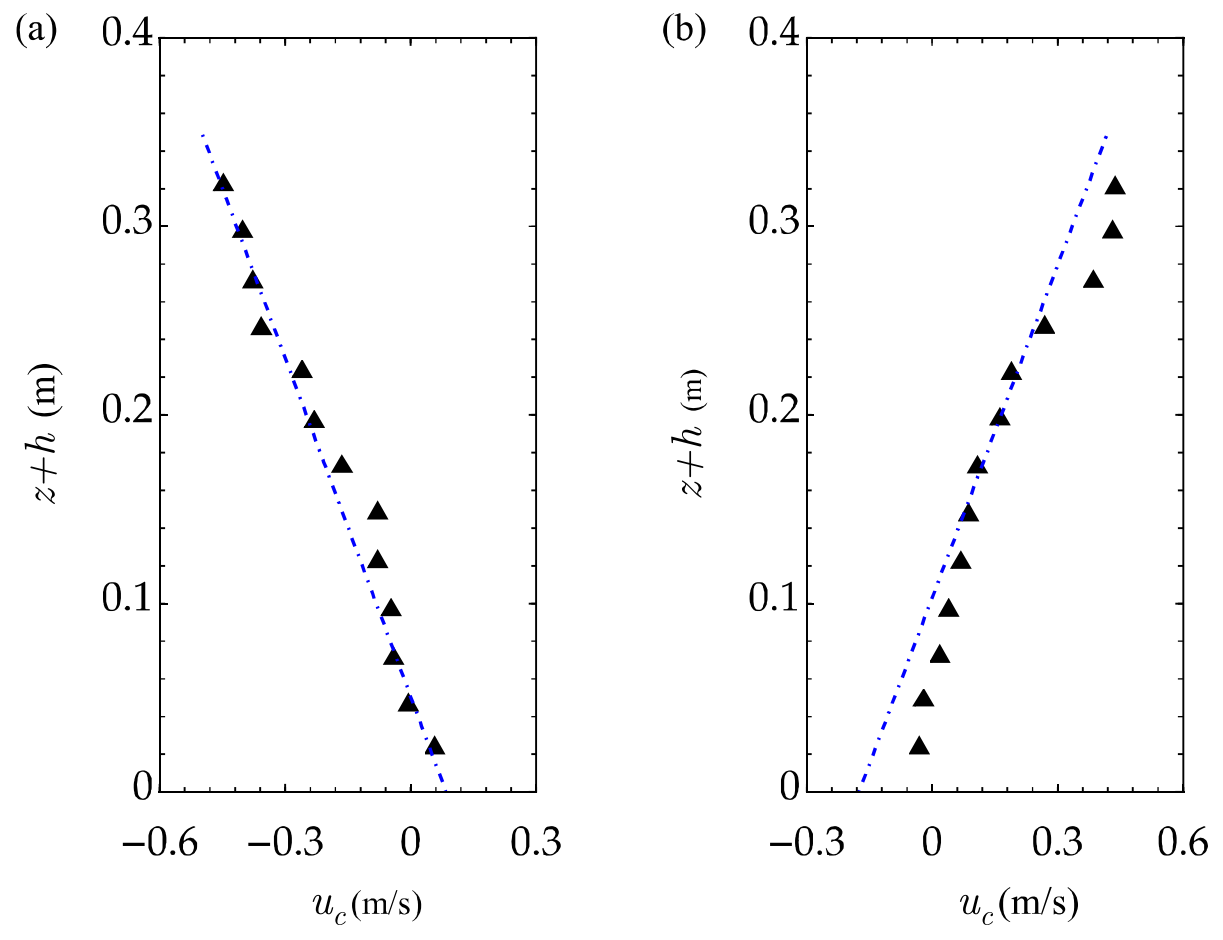

**Figure 5.** Comparisons between the current profile employed in the present solutions (blue dash-dotted line) and the measured current data by Swan [25] (black triangle) for (a) case S1; (b) case S2.

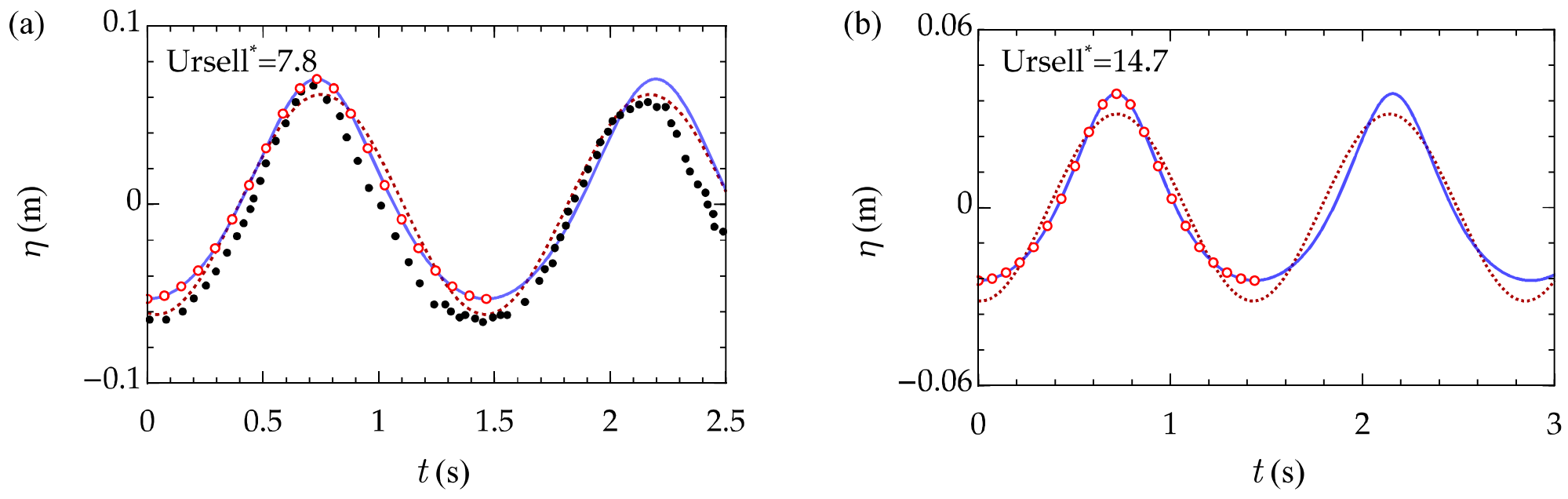

**Figure 6.** Comparisons of the surface elevation $(x = \pi/k)$ among the present fifth-order solutions (blue solid line), experimental data of Swan [25] (black dot), numerical solutions of Francius & Kharif [7] (red circle), as well as the linear analytical solutions of Touboul et al. [26] (red dotted line) for (a) case S1; (b) case S2.

In figure 7, the comparisons of the horizontal velocity profiles under the wave crest among various solutions are depicted. Satisfactory agreements between the present solution and the numerical solutions are obtained, despite the discrepancies with the measured data of Swan [25]. One possible reason for the discrepancies is linked to the bias of the linear current profile approximated from the experimental data, as presented in figure 5 (b), where, near the bottom, the approximated current profile is slightly smaller than the measurements, thereby leading to the underestimation in the velocity profile for case S2.

As shown in figures 7 (a) and 7 (b), with the enhancement of wave nonlinearity, namely, Ursell* value increases from 7.8 to 14.7, the present solutions provide better comparisons than the linear analytical solutions of Touboul et al. [26], indicating the capacities of the proposed solutions in depicting Stokes waves with strong nonlinearity. Moreover, the present solutions agree well with the numerical solutions, in which the same linearized current profile was applied in the calculation, demonstrating the prediction capability of the present fifth-order solution for Stokes wave propagation with currents. The comparable wave speed and horizontal velocities suggest that the present solutions can predict the Stokes waves under both strong and weak linear shear current conditions comparatively well and the solutions have the potential to be extended to a wider range of applications.

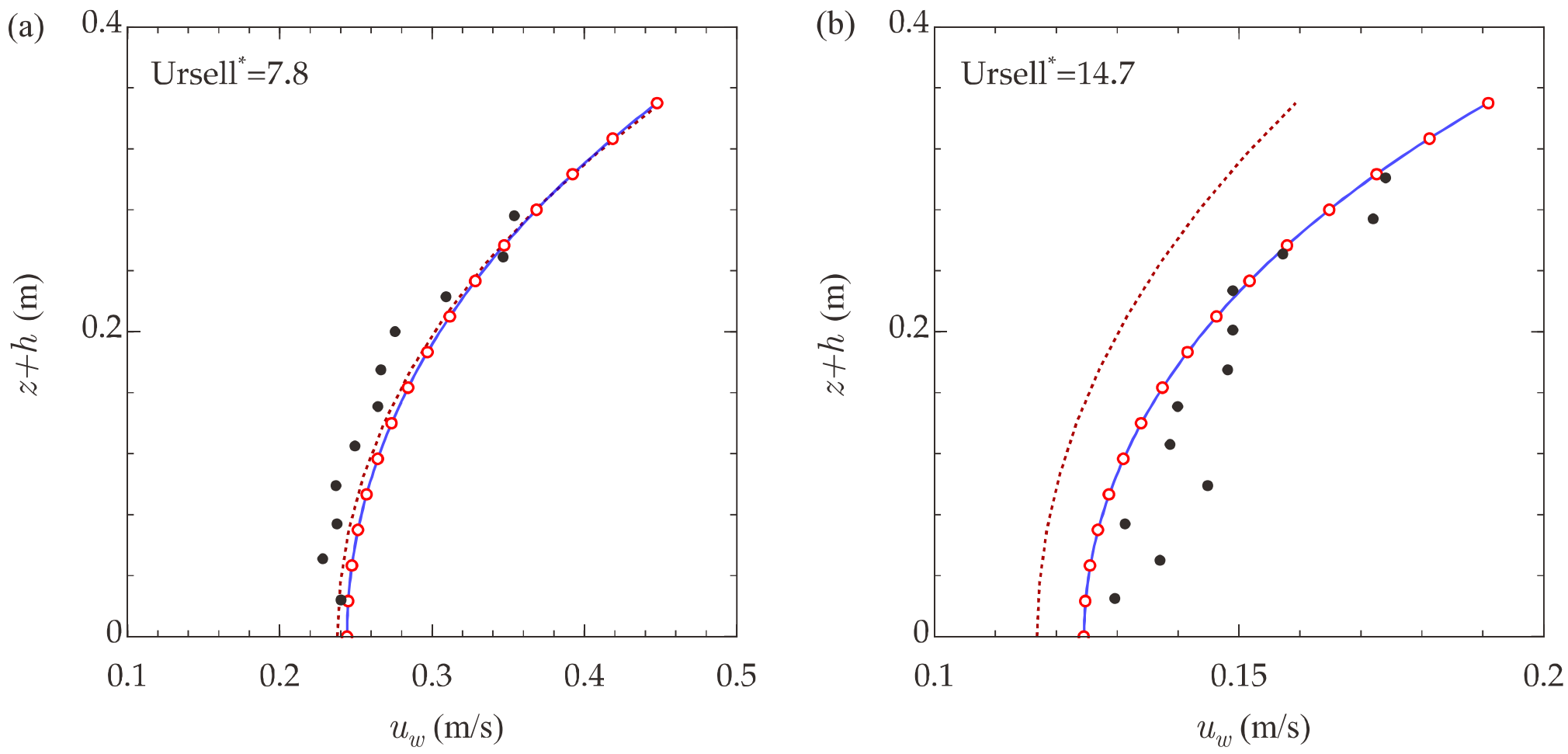

**Figure 7.** Comparisons of the horizontal velocity under the wave crest among the present fifth-order solutions (blue solid line), experimental data of Swan [25] (black dot), numerical solutions of Francius & Kharif [7] (red circle), as well as the linear analytical solutions of Touboul et al. [26] (red dotted line) for (a) case S1; (b) case S2.

## 4. Solution performance in describing wave motion and dynamic properties

The present solutions for describing Stokes wave propagation in linear shear currents have been validated by comparing with existing experimental and numerical results. In this section, the effects of wave nonlinearity, averaged current and vorticity on the wave motion and dynamic properties are examined by using the present solutions, together with the third-order Stokes wave solutions proposed by Kishida & Sobey [22], for further insight discussions. The work of Kishida & Sobey [22] considered a linear shear current and is the counterpart to Fenton's solution [12] for the pure wave condition. Both solutions were based on the stream function approach, and adopted the non-physical assumption, $H = 2A$, as discussed in the previous section. Finally, the properties of the fluid particle trajectories are investigated under different linear shear current conditions by using the present solutions.

### (a) Stokes wave solutions under different wave nonlinearity conditions

In this section, the influences of wave nonlinearity on surface elevation, wave speed and velocity distribution are investigated by for two sets of wave and current conditions as shown in table 3. All the parameters are the same for case (I) and (II) except for the water depth, which is set to $0.23\,\mathrm{m}$ and $0.20\,\mathrm{m}$, respectively, resulting in $H/h = 0.27$ and $0.32$ and Ursell $^* = 36$ and $49$, respectively.

Figure 8 illustrates the comparisons of the surface elevation and horizontal velocity profile in the water column between the present solutions and other existing results. The wave surface elevations and horizontal velocity profiles predicted by the present fifth-order solutions compare well with the numerical solutions, indicating that the present Stokes solutions can well describe waves with strong nonlinearity. When Ursell$^* = 36$, the third-order solutions by Kishida & Sobey [22] show fluctuations along the wave trough, resulting in more obvious discrepancies for the velocity profiles in the water column, which demonstrate the importance of the higher-order components in the solutions. For the case of shallower water depth, namely, the Ursell$^* = 49$, the differences between the present solutions and the third-order solutions of Kishida & Sobey [22] increase, particularly in the wave trough, resulting in more noticeable differences in vertical velocity profiles. These significant differences are linked to the sharp decrease in the denominator of $\gamma$ in equation (2.30), which brings great challenges for the third-order solution of Kishida & Sobey [22] to be effective with the substantial increase in Ursell$^*$.

**Table 3.** Parameters for cases with different water depth.

| case | $H$(mm) | $h$(m) | $T$(s) | $\bar{U}$(m/s) | $\Omega$ (/s) | $H/h$ | $kh$ | $U'$ | $\Omega'$ | Ursell* |
|---|---|---|---|---|---|---|---|---|---|---|
| (I) | 63 | 0.23 | 1.418 | 0.123 | 1.7 | 0.27 | 0.63 | 0.08 | 0.38 | 36 |
| (II) | 63 | 0.20 | 1.418 | 0.123 | 1.7 | 0.32 | 0.57 | 0.08 | 0.38 | 49 |

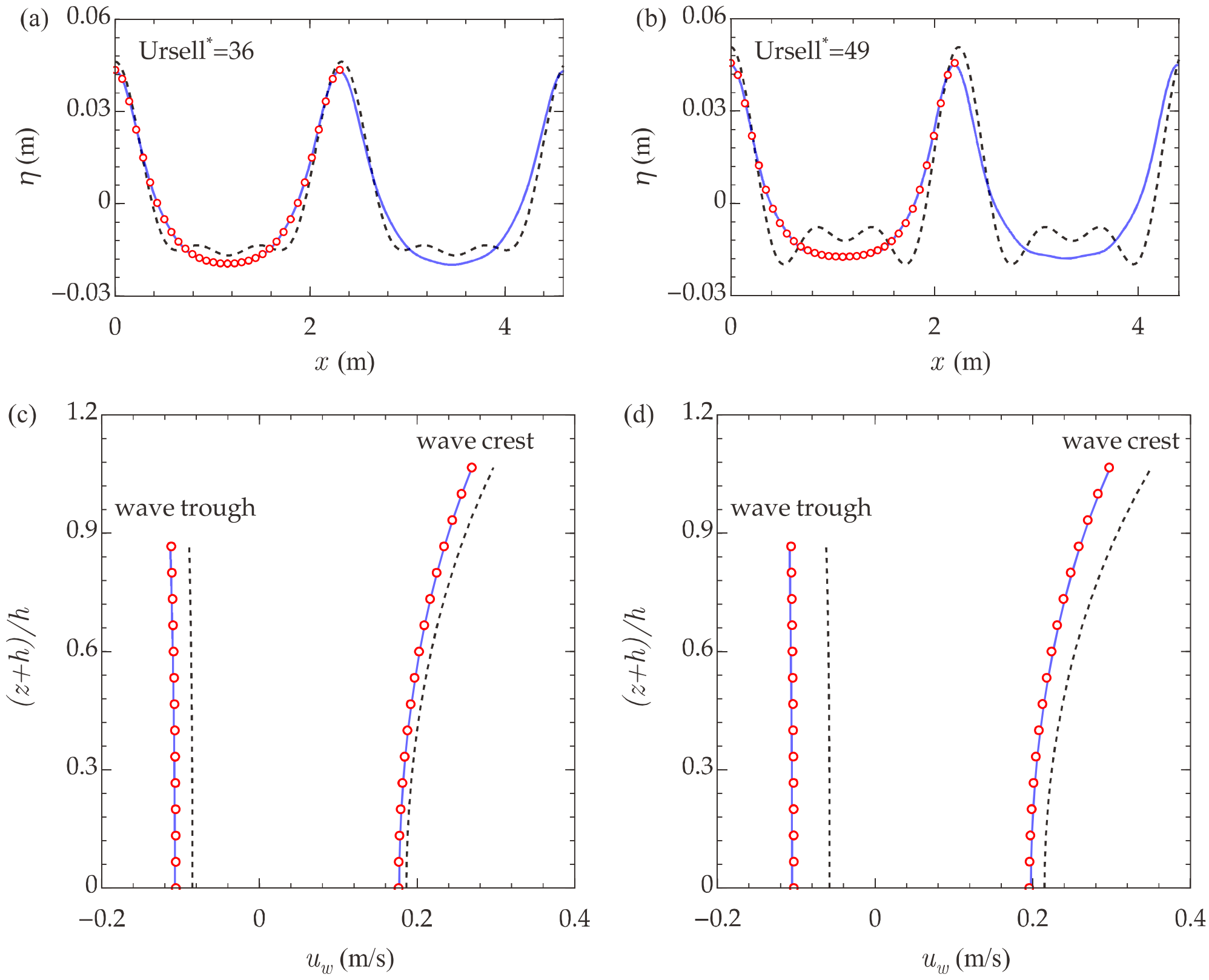

**Figure 8.** Comparisons of surface elevations $(t=0)$ and horizontal velocity profiles in the water column under the wave crest and wave trough for different water depths among the present fifth-order solutions (blue solid line), the numerical solutions of Francius & Kharif [7] (red circle) and the third-order solutions of Kishida & Sobey [22] (black dashed line). Panels (a) and (c) are for case (I) and panels (b) and (d) are for case (II).

As illustrated in figure 9, with the further increase of $H/h$, the discrepancies in the predicted phase speed become more pronounced for the solutions of Kishida & Sobey [22]. The present solutions remain effective for estimating the phase speed of highly nonlinear waves, attributing to the closely coupled relation between dispersion relation and wave motion, in which the dispersion equation (2.27), the wave height equation (2.28) and surface elevation (2.26) are solved simultaneously and therefore guarantee a strong restriction to wave motion, providing accurate predictions of phase speed, velocity distribution and wave elevation even under the strong nonlinearity conditions. While in the derivation of Kishida & Sobey [22], these strongly coupled relations are simplified by assuming $H=2A$, and only the dispersion equation is solved. As a result, the third-order solution of Kishida & Sobey [22] shows inaccurate prediction of the phase speed that cannot perfectly match the real wave motion, thereby resulting in perturbations in the waves elevation and velocity distributions, especially for waves with strong nonlinearity.

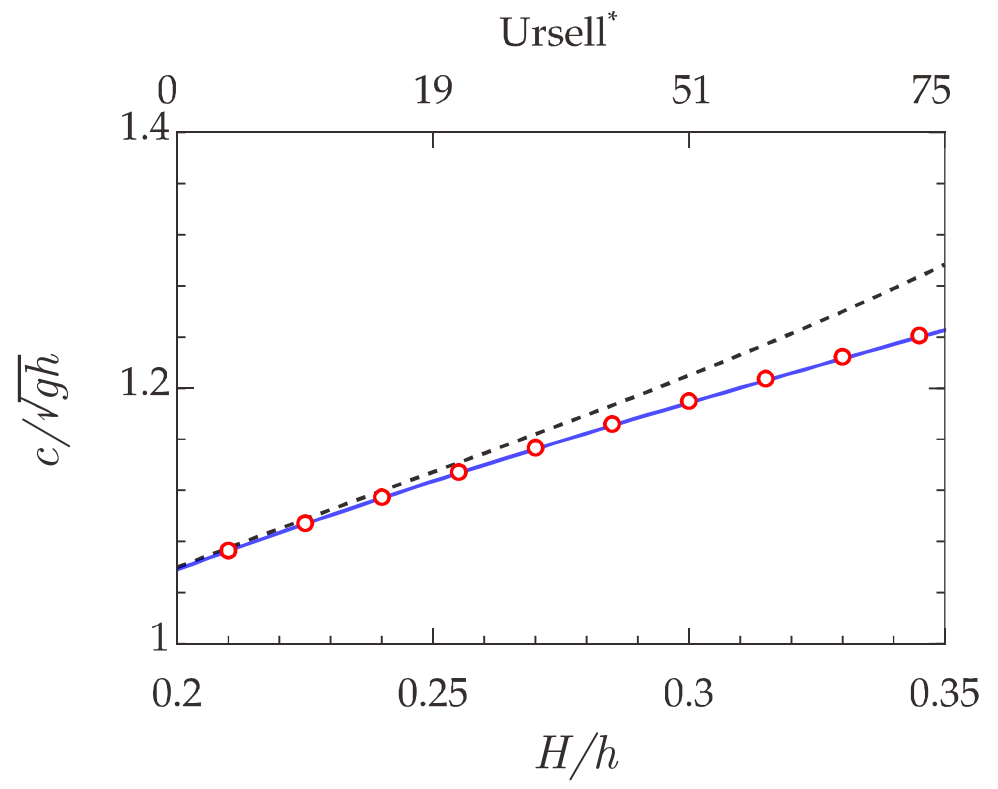

**Figure 9.** The normalized phase velocity $c/\sqrt{gh}$ varies with $H/h$ among the present fifthorder solution (blue solid line), the numerical solutions of Francius & Kharif [7] (red circle) and third-order solutions of Kishida & Sobey [22] (black dashed line).

## (b) Stokes wave solutions under different averaged current conditions

Further investigations are conducted to analyze the present solutions' performance under two different averaged current conditions. As shown in table 4, the wave parameters and the vorticity are the same for the both cases. However, except the averaged current, $\bar{U}$, is set to $-0.245$ m/s and $0.735$ m/s for case (I) and (II), respectively.

**Table 4.** Parameters for cases with different average current.

| case | $H$(mm) | $h$(m) | $T$(s) | $\bar{U}$(m/s) | $\Omega$ (/s) | $H/h$ | $kh$ | $U'$ | $\Omega'$ | Ursell* |
|---|---|---|---|---|---|---|---|---|---|---|
| (I) | 94.5 | 0.35 | 1.418 | -0.245 | 1.7 | 0.27 | 1.03 | -0.16 | 0.38 | 14 |
| (II) | 94.5 | 0.35 | 1.418 | 0.735 | 1.7 | 0.27 | 0.59 | 0.28 | 0.38 | 42 |

Figure 10 depicts the surface elevations and horizontal velocity profiles in the water column for different $\bar{U}$. Figures 10 (a) and 10 (c) illustrate the surface elevations and velocity profiles for case (I), in which the averaged current is in the opposite direction of wave propagation. The present fifth-order solution produces satisfactory results as compared with the numerical results. On the other hand, the third-order solutions [22] show slight differences for the surface elevation and velocity profiles near the free surface, demonstrating the significant importance of higher-order terms for Stokes wave solutions. Figures 10 (b) and 10 (d) show the surface elevation and velocity profiles for case (II), where the waves propagate in the same direction as the averaged current and the Ursell* exhibits a significant increase, due to the sharp decrease in the denominator of $\gamma$ caused by the positive averaged current ($\bar{U}>0$), as illustrated in equations (2.29) and (2.30). The third-order solutions of Kishida & Sobey [22] show noticeably secondary fluctuations along the wave trough, resulting in significant discrepancy in the velocity profiles in the water column. And the agreement between the present solutions and the numerical solutions further verifies the robustness of the present solutions for Stokes waves in strong averaged currents, even when Ursell* exceeds 40.

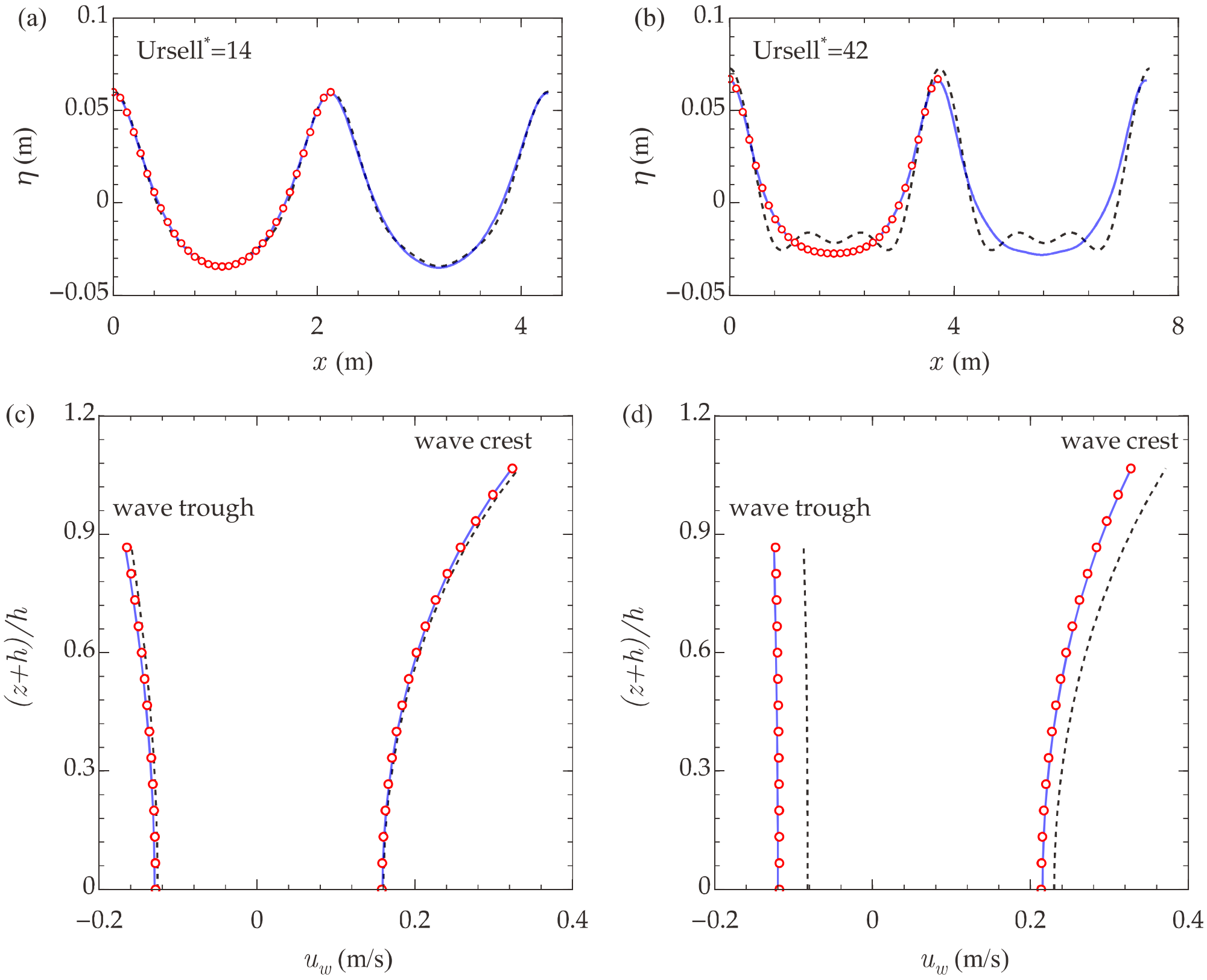

**Figure 10.** Comparisons of surface elevations $(t = 0)$ and horizontal velocity profiles in the water column under wave crest and wave trough for different averaged current among the present fifth-order solutions (blue solid line), the numerical solutions of Francius & Kharif [7] (red circle) and the third-order solutions of Kishida & Sobey [22] (black dashed line). Panels (a) and (c) for case (I); panels (b) and (d) for case (II).

## (c) Stokes wave solutions under different vorticity conditions

In this section, the effects of vorticity on the surface elevation and velocity profile are investigated. Two vorticity strengths are employed, i.e., $-1.7$/s and $2.7$/s for case (I) and (II), respectively. The other wave and current parameters are set the same, which are listed in table 5.

**Table 5.** Parameters for cases with different vorticity.

| case | $H$(mm) | $h$(m) | $T$(s) | $\bar{U}$(m/s) | $\Omega$ (/s) | $H/h$ | $kh$ | $U'$ | $\Omega'$ | Ursell* |
|---|---|---|---|---|---|---|---|---|---|---|
| (I) | 63 | 0.25 | 1.418 | 0.123 | -1.70 | 0.25 | 0.64 | 0.08 | -0.38 | 16 |
| (II) | 63 | 0.25 | 1.418 | 0.123 | 2.72 | 0.25 | 0.64 | 0.07 | 0.61 | 38 |

Figure 11 illustrates the comparisons of wave surface elevation and velocity profiles for Stokes waves propagation under different vorticity conditions. As illustrated in equations (2.29) and (2.30), the direction of the vorticity has a notable impact on the magnitude of Ursell* since the symbol of $\gamma$ is closely related to the vorticity $\Omega$. A negative vorticity $(\Omega < 0)$ results in $\gamma < 0$, causing a significant decrease in Ursell*. Conversely, a positive vorticity $(\Omega > 0)$ leads to a notable increase in the magnitude of Ursell*, posing challenges for low-order solutions to accurately capture the behavior of waves propagation. Figures 11 (a) and 11 (c) show satisfactory agreement in surface elevation and velocity profile among all the solutions when the vorticity is negative, with Ursell* $= 16$. For the waves propagating in positive vorticity, where Ursell* is much greater than that under negative vorticity condition, the third-order solutions of Kishida & Sobey [22] fail to provide accurate wave surface elevation and velocity profiles in the water column, as depicted in figures 11 (b) and 11 (d). This is not only because the high-order effect is not considered in the third-order solution of Kishida & Sobey [22], but also due to the absence of a strong connection between surface elevation and dispersion relation, which arises from the non-physical prior hypothesis employed in their study, $H = 2A$. The present fifth-order solutions remain valid in high vorticity scenarios. This further demonstrates their superior applicability in predicting wave propagation in strong shear currents.

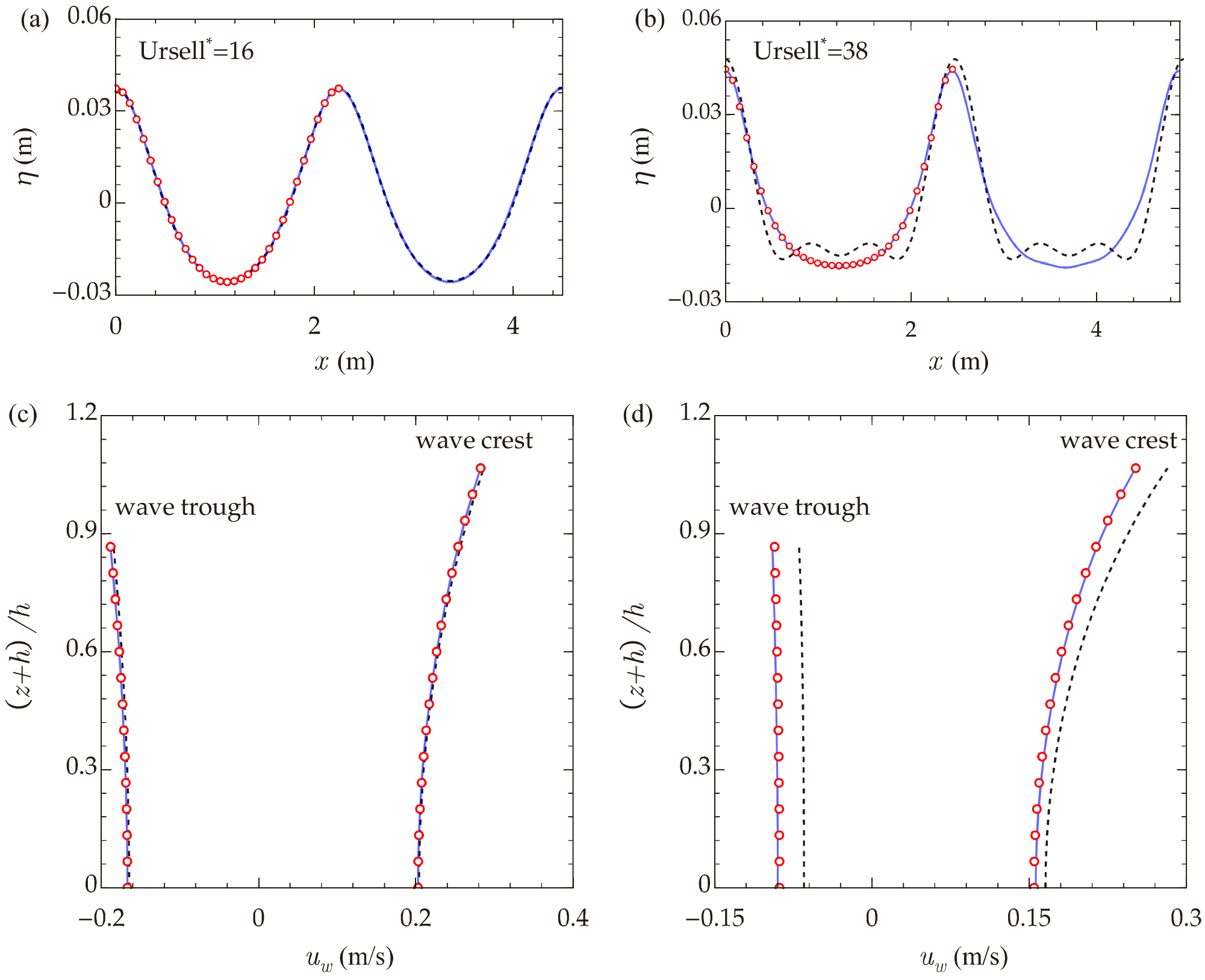

**Figure 11.** Comparisons of surface elevations $(t = 0)$ and velocity profiles in the water column under wave crest and wave trough for different vorticity among the present fifth-order solutions (blue solid line), the numerical solutions of Francius & Kharif [7] (red circle) and the third-order solutions of Kishida & Sobey [22] (black dashed line). Panels (a) and (c) for case (I); panels (b) and (d) for case (II).

## (d) Fluid particle trajectories

In this section, the fluid particle trajectories in various linear shear currents are investigated. The concept of Stokes drift, initially studied by Stokes [10], refers to the forward drift of fluid particles in the direction of wave propagation after one wave period. The presence of a linear shear current alterates the fluid particle trajectories, where the fluid particles can move both forwards and backwards rather than moving along a fixed direction, which has been examined in previous studies [24,29–32].

The coordinates of the fluid particle's motion, denoted by $x = x(t)$ and $z = z(t)$, can be obtained by integrating the following differential equations

$$\frac{dx}{dt} = \phi_x + U_0 + \Omega z, \quad \frac{dz}{dt} = \phi_z, \tag{4.1}$$

in which the velocity potential function $\phi$ is defined in equation (2.24). The fluid particle trajectories under four different current conditions, namely, the pure wave for case (I), a positive vorticity for case (II), a negative vorticity for case (III) and a surface current combined positive vorticity for case (IV), are analyzed. The wave and current parameters are listed in table 6.

**Table 6.** Parameters for cases under different current condtions.

| case | $H$(m) | $h$(m) | $k$(/m) | $T$(s) | $U_0$(m/s) | $\Omega$ (/s) |
|---|---|---|---|---|---|---|
| (I) | 0.1 | 0.5 | 1 | 2.9 | 0 | 0 |
| (II) | 0.1 | 0.5 | 1 | 3.0 | 0 | 0.5 |
| (III) | 0.1 | 0.5 | 1 | 2.7 | 0 | -0.5 |
| (IV) | 0.1 | 0.5 | 1 | 2.9 | 0.1 | 0.5 |

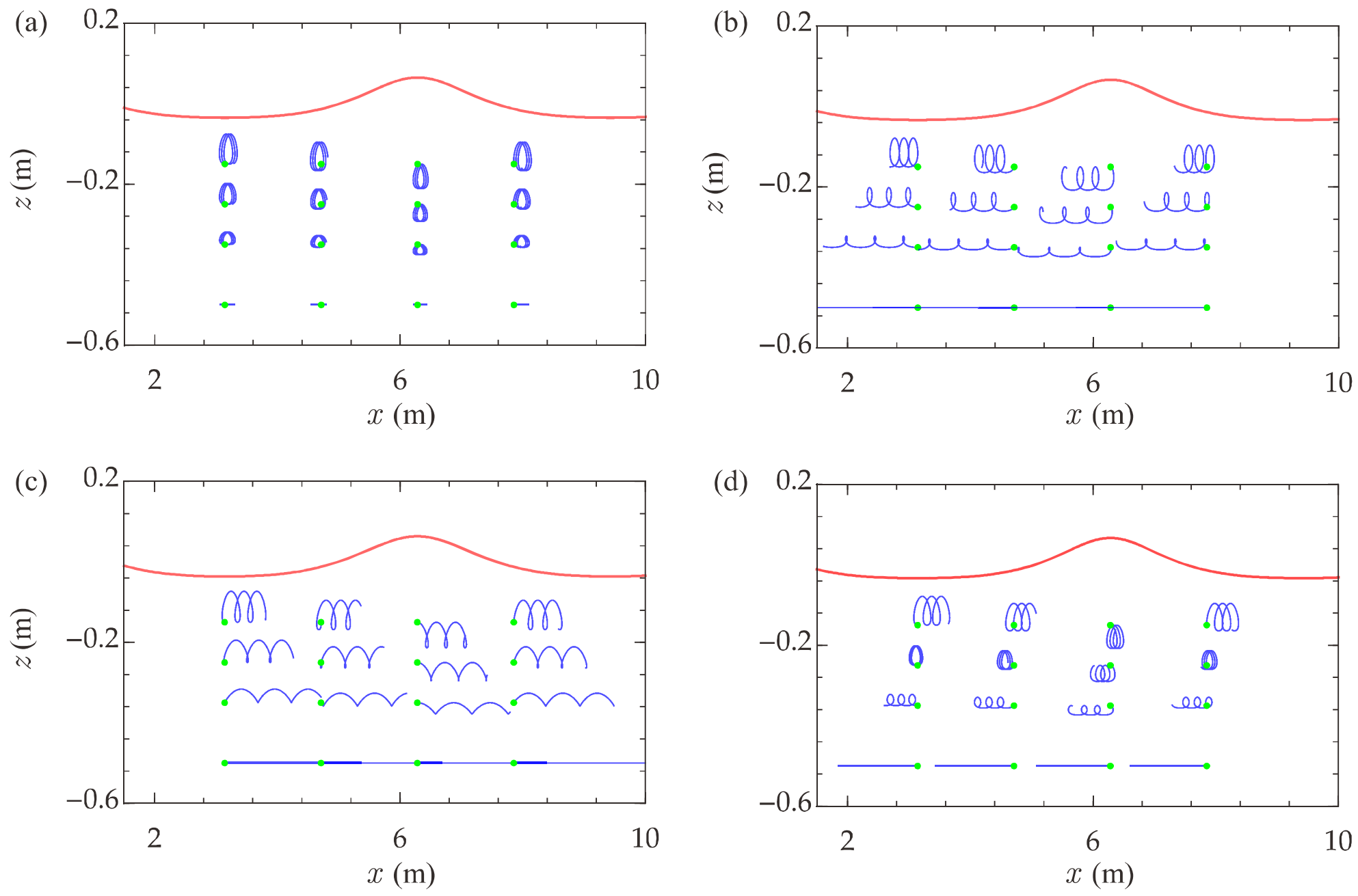

**Figure 12.** Fluid particle trajectories (blue line) over three wave periods $(0 \sim 3T)$ under different current conditions, where surface elevation at the initial moment $(t = 0)$ is represented by red line and the initial positions of particles are represented by green dots. There are total 12 fluid particles whose horizontal positions are $x = (\pi/k, 1.5\pi/k, 2\pi/k$ and $2.5\pi/k)$, with the vertical coordinates $z = (-0.15\ \text{m}, -0.25\ \text{m}, -0.35\ \text{m}$ and $-0.5\ \text{m})$, respectively. Panel (a) for case (I); panel (b) case (II); panel (c) for case (III); panel (d) for case (IV).

Figure 12 (a) illustrates the periodic forward drift of fluid particles under a pure wave condition, where the movements of fluid particles are either spiral-shaped (for those near the surface) or line-shaped (for those near the bottom). When the background vorticity exists, it can significantly alter the direction and pattern of fluid particle motions as shown in figures 12 (b) and 12 (c). While the positive shear results in backward particle drift, where the motion of fluid particles is dominant by the effect of shear, as displayed in figure 12 (b), the negative shear intensifies the forward drift due to the vorticity-induced forward current (figure 12 (c)). Note that in cases (II) and (III) the surface current is zero and the direction of fluid particle drifts remains the same in the water column, depending on the sign of the background vorticity. In case (IV), the surface current has a positive value, figure 12 (d) shows that the direction of the fluid particle drifts changes from positive near the free surface to negative near the bottom. These distinct behaviors are led by the combined effects of surface current, vorticity, and waves. These three geometric shapes of fluid trajectories, in figures 12 (b)-(d), are consistent with the observations in [32], which further highlights the effects of linear shear currents on changing the geometry of the trajectories.

## 5. Conclusion

In this paper, a new set of fifth-order Stokes solutions for periodic waves in a linear shear current is derived. This work is an extension of the fifth-order Stokes wave solutions without current (Zhao & Liu [15]) with the use of the same perturbation techniques. The proposed solutions are first checked with the experimental data of comparatively weak currents by Thomas [28] and Swan [25] and the numerical solutions of Francius & Kharif [7] using the spectral method. While excellent agreement between the present solutions and numerical solutions is observed, slight differences in velocity profiles in the water column between the present solutions and experimental data are observed due to the discrepancy between the measured current data and the approximated linear current profile used in the present solutions.

Subsequently, the performance of the present solutions for under the conditions of stronger wave nonlinearity, larger mean current speed, and larger vorticity, the performance of the present solutions is examined, together with the third-order Stokes wave solutions proposed by Kishida & Sobey [22]. The superiority of the present solutions in predicting highly nonlinear Stokes waves and their propagation in a positive averaged (following) current and vorticity is demonstrated. The relatively unsatisfactory performance of the theoretical solutions by Kishida & Sobey [22] is caused by the adoption of the non-physical assumption of $H = 2A$ (wave height must be the twice of the wave amplitude of the first-order first harmonic wave), which is not ensured when wave nonlinearity is strong. Instead, in the present solutions the wave height is part of the solutions by jointly solving the dispersion equation, the wave height equation and the surface elevation. Therefore, the present solution guarantees accurate predictions of wave speed, surface elevation and velocity distribution even under very large Ursell$^*$ conditions.

Furthermore, based on the proposed solutions, the multiple moving modes of fluid particle trajectories are observed. The results reveal that the vorticity can alter the direction of particle motion, resulting in the intensity of forward movements for negative shear or backward particle trajectories for positive shear current . In addition, the incorporation of surface current can lead to reverse motion behaviors for vertically distributed particles, and both positive and negative modes of the particles that are under the wave crest have been observed.

Data Accessibility. Additional data are available in the electronic supplementary material, and the software resource is available at https://github.com/474278604/Haiqi-Fang.git.

Authors' Contributions. H.F.: formal analysis, methodology, validation, writing - original draft; P.L.-F.L.: conceptualization, supervision, writing - review & editing; L.T.: formal analysis, investigation, writing - review & editing; P.L.: conceptualization, supervision.

Competing Interests. We declare we have no competing interests.

Acknowledgements. This work is supported, in part, by the National Natural Science Foundation of China (Grant No. 52031002); Open Funding of State Key Laboratory of Coastal and Offshore Engineering, Dalian University of Technology (Grant No. LP2004); the Open Funding of Key Laboratory of Coastal Disaster and Protection of Hohai University (Grant No. 202213). This research is also supported in part by the Yushan Program, Ministry of Education in Taiwan.

# SUPPLEMENTARY MATERIAL FOR RSPA

## Supporting Information for

### The theory of fifth-order Stokes waves in a linear shear current

**Haiqi Fang, Philip L.-F. Liu, Lian Tang, Pengzhi Lin**

**This PDF file includes:**

Supporting text
SI References

## Supporting Information Text

### 1. Procedure to derive the fifth-order solutions

The governing equation and boundary conditions for the velocity potential can be specified as

$$\phi_{xx} + \phi_{zz} = 0. \quad -h \le z \le \eta \tag{1}$$

At the sea bottom, the no-flux boundary condition is

$$\phi_z = 0. \quad z = -h \tag{2}$$

The Taylor expansions of the combined free surface and dynamic boundary conditions at $z = 0$ yield:

$$\sum_{n=0}^{\infty} \frac{\eta^n}{n!} \frac{\partial^n}{\partial z^n} \left\{ \beta^2 \phi_{\tau\tau} + g\phi_z + \left( \beta \frac{\partial}{\partial \tau} + (U_0 + \Omega\eta + \phi_x) \frac{\partial}{\partial x} + \phi_z \frac{\partial}{\partial z} \right) \times \left[ \beta\phi_\tau + \frac{1}{2}\left(\phi_x^2 + \phi_z^2\right) + (U_0 + \Omega\eta)\,\phi_x - \Omega\psi \right] - \beta^2 \phi_{\tau\tau} \right\} = 0, \quad z = 0 \tag{3}$$

and

$$\eta = -\frac{1}{g} \sum_{n=0}^{\infty} \frac{\eta^n}{n!} \frac{\partial^n}{\partial z^n} \left\{ \beta\phi_\tau + (U_0 + \Omega\eta)\,\phi_x + \frac{1}{2}\left(\phi_x^2 + \phi_z^2\right) - \Omega\psi \right\}, \quad z = 0 \tag{4}$$

in which a stretched time variable $\tau$ is used to replace the original time variable $t$ :

$$\tau = \beta t, \tag{5}$$

where

$$\beta = 1 + \sum_{i=1}^{\infty} \varepsilon^{(2\times i)} \beta_{(2\times i)}. \tag{6}$$

We seek for the following perturbation series forms of Stokes wave solutions, written as:

$$\phi = \sum_{i=1}^{\infty} \phi_i \varepsilon^i, \tag{7}$$

$$\psi = \sum_{i=1}^{\infty} \psi_i \varepsilon^i, \tag{8}$$

$$\eta = \sum_{i=1}^{\infty} \eta_i \varepsilon^i. \tag{9}$$

Substituting the perturbation series solutions, equations (7)-(9), into the Laplace equation, (1), as well as the boundary conditions, (2)-(4), and collecting the terms with the same order in terms of $\varepsilon^i$, the $i$-order governing equation and corresponding boundary conditions are derived. For the $i$-order problem, the governing equation as well as the bottom boundary condition are given by $\Delta\phi_i = 0$, $\partial\phi_i/\partial z = 0$, respectively, providing the general solutions for $\phi_i$.

$$\phi_i = C_i\tau + \sum_{j=1}^{N_i} A_{ij} \cosh k_j(z+h) \sin\left(k_j x - \omega_j \tau\right), \tag{10}$$

and $\psi_i$ can be obtained

$$\psi_i = \sum_{j=1}^{N_i} A_{ij} \sinh k_j(z+h) \cos\left(k_j x - \omega_j \tau\right). \tag{11}$$

where $C_i, A_{ij}, k_j, \omega_j$ are coefficients and the number of terms for summation, $N_i$, will be determined through the other two boundary conditions at $z = 0$.

**A. First-order equations and solutions.** By collecting the coefficients with $\varepsilon^1$ in (3), the following combined boundary condition is obtained.

$$\frac{\partial^2\phi_1}{\partial\tau^2}+g\frac{\partial\phi_1}{\partial z}+U_0^2\frac{\partial^2\phi_1}{\partial x^2}+2U_0\frac{\partial^2\phi_1}{\partial x\partial\tau}-\Omega\left(\frac{\partial\psi_1}{\partial\tau}+U_0\frac{\partial\psi_1}{\partial x}\right)=0, z=0 \quad [12]$$

Similarly, (4) gives the dynamic boundary condition for $\eta_1$.

$$\eta_1=-\frac{1}{g}\left(\frac{\partial\phi_1}{\partial\tau}+U_0\frac{\partial\phi_1}{\partial x}-\Omega\psi_1\right). \quad [13]$$

By assuming the first order wave is a monochromatic wave with wavenumber $k$, wave frequency $\omega_0$, and wave amplitude $a$. The transform (5) and (6) , the settings of the first order problem are determined as $C_1=0$, $N_1=1, k_1=k, \omega_1=\omega_0$. Therefore, the first order solutions are proposed as:

$$\phi_1=\frac{gA\sqrt{1-\sigma^2}}{\rho}\cosh k(z+h)\sin\left(kx-\omega_0\tau\right), \quad [14]$$

$$\psi_1=\frac{gA\sqrt{1-\sigma^2}}{\rho}\sinh k(z+h)\cos\left(kx-\omega_0\tau\right), \quad [15]$$

$$\eta_1=A\cos\left(kx-\omega_0\tau\right), \quad [16]$$

in which

$$\sigma=\tanh kh, \quad [17]$$

$$\rho=\omega_0-kU_0+\Omega\tanh kh. \quad [18]$$

**B. Second-order equations and solutions.** For the second order problem, the combined and the dynamic free surface conditions applied at $z=0$ are given as follows:

$$\begin{aligned}&\frac{\partial^2\phi_2}{\partial\tau^2}+g\frac{\partial\phi_2}{\partial z}+U_0^2\frac{\partial^2\phi_2}{\partial x^2}+2U_0\frac{\partial^2\phi_2}{\partial x\partial\tau}-\Omega\left(\frac{\partial\psi_2}{\partial\tau}+U_0\frac{\partial\psi_2}{\partial x}\right)\\
&=-\frac{\partial^2\phi_1}{\partial z\partial\tau}\frac{\partial\eta_1}{\partial\tau}-\frac{\partial^2\phi_1}{\partial x\partial\tau}\frac{\partial\phi_1}{\partial x}-\frac{\partial^2\phi_1}{\partial z\partial\tau}\frac{\partial\phi_1}{\partial z}+g\frac{\partial\eta_1}{\partial x}\frac{\partial\phi_1}{\partial x}-\Omega\frac{\partial\eta_1}{\partial\tau}\frac{\partial\phi_1}{\partial x}\\
&\quad+\Omega\frac{\partial\eta_1}{\partial\tau}\frac{\partial\psi_1}{\partial z}-\eta_1\left\{\Omega\left(\frac{\partial^2\phi_1}{\partial x\partial\tau}-\frac{\partial^2\psi_1}{\partial z\partial\tau}-g\frac{\partial\eta_1}{\partial x}\right)+\frac{\partial^3\phi_1}{\partial x\partial\tau^2}+g\frac{\partial^2\phi_1}{\partial z^2}\right\}\\
&\quad-U_0\left\{\frac{\partial^2\phi_1}{\partial z\partial\tau}\frac{\partial\eta_1}{\partial x}+\frac{\partial^2\phi_1}{\partial x\partial z}\frac{\partial\eta_1}{\partial\tau}+\frac{\partial^2\phi_1}{\partial x\partial z}\frac{\partial\phi_1}{\partial z}\right.\\
&\quad\left.+\Omega\left[\frac{\partial\eta_1}{\partial x}\left(\frac{\partial\phi_1}{\partial x}-\frac{\partial\psi_1}{\partial z}\right)-\frac{\partial^2\psi_1}{\partial x\partial z}\eta_1\right]+2\eta_1\frac{\partial^3\phi_1}{\partial x\partial z\partial\tau}+\left(\frac{\partial\phi_1}{\partial x}+\Omega\eta_1\right)\frac{\partial^2\phi_1}{\partial x^2}\right\}\\
&\quad-U_0^2\left(\frac{\partial^2\phi_1}{\partial x\partial z}\frac{\partial\eta_1}{\partial x}+\eta_1\frac{\partial^3\phi_1}{\partial x^2\partial z}\right),\end{aligned} \quad [19]$$

$$\eta_2=-\frac{1}{2g}\left\{\left(\frac{\partial\phi_1}{\partial x}\right)^2+\left(\frac{\partial\phi_1}{\partial z}\right)^2+2\eta_1\left(\frac{\partial^2\phi_1}{\partial z\partial\tau}+\Omega\frac{\partial\phi_1}{\partial x}-\Omega\frac{\partial\psi_1}{\partial z}+U_0\frac{\partial^2\phi_1}{\partial x\partial z}\right)-2\Omega\psi_2+2\frac{\partial\phi_2}{\partial\tau}+2U_0\frac{\partial\phi_2}{\partial x}\right\}. \quad [20]$$

On the right-hand sides of the equations above are terms determined by the first order solutions $\phi_1, \psi_1$, $\eta_1$ and their derivatives at $z=0$.

The corresponding second-order solutions can be expressed as:

$$\phi_2=C_2\tau+\frac{A^2\left(1-\sigma^2\right)\left[\rho^2+\rho\chi+\left(1-3\sigma^2\right)\chi^2\right]}{8\chi\sigma^4}\cosh 2k(z+h)\sin\left(2kx-2\omega_0\tau\right), \quad [21]$$

$$\psi_2=\frac{A^2\left(1-\sigma^2\right)\left[\rho^2+\rho\chi+\left(1-3\sigma^2\right)\chi^2\right]}{8\chi\sigma^4}\sinh 2k(z+h)\cos\left(2kx-2\omega_0\tau\right), \quad [22]$$

$$\eta_2=\frac{kA^2\left[\rho^2+\rho\left(1+\sigma^2\right)\chi+\left(1-2\sigma^2\right)\chi^2\right]}{4\sigma^3\chi^2}\cos\left(2kx-2\omega_0\tau\right)+\frac{\chi^2A^2\left(-1+\sigma^2\right)}{4g\sigma^2}-\frac{C_2}{g}, \quad [23]$$

in which

$$\chi=\omega_0-kU_0. \quad [24]$$

Similar to the derivation of Zhao & Liu (1), the coordinate system is set on the still water level with $h$ being the water depth, resulting in zero mean free surface, i.e., $\overline{\eta_2}=0$ . Therefore, the constant term in equation (23) is equal to 0, yielding

$$C_2=\frac{\chi^2A^2\left(-1+\sigma^2\right)}{4\sigma^2}. \quad [25]$$

**C. Third-order equations and solutions.** For the third-order problem, the combined and the dynamic free surface boundary conditions applied at $z = 0$ are given as follows:

$$
\begin{aligned}
&\frac{\partial^2\phi_3}{\partial\tau^2} + g\frac{\partial\phi_3}{\partial z} + U_0^2\frac{\partial^2\phi_3}{\partial x^2} + 2U_0\frac{\partial^2\phi_3}{\partial x\partial\tau} - \Omega\left(\frac{\partial\psi_3}{\partial\tau} + U_0\frac{\partial\psi_3}{\partial x}\right)\\
&\quad = \left\{ -\frac{kA^3\left[\rho^4 + 2\rho^3\left(1+\sigma^2\right)\chi - 3\rho^2\left(-1+\sigma^2\right)\chi^2 + 2\rho\left(-1+\sigma^2\right)^2\chi^3 + \left(1-5\sigma^2+7\sigma^4\right)\chi^4\right]}{8\sigma^5\chi} \right.\\
&\quad \left. + \frac{A\omega_0\chi(\rho+\chi)}{k\sigma}\beta_2 \right\}\sin\left(kx - \omega_0\tau\right)\\
&\quad - \frac{3A^3k}{8\rho\sigma^4\chi^2}\left\{-4\rho\sigma\chi^3\left[\rho^2 + \rho\chi + \left(1-3\sigma^2\right)\chi^2\right] + g^2k^2\sigma\left[\rho^2 + \rho\left(1+\sigma^2\right)\chi + \left(-1+\sigma^2\right)^2\chi^2\right] + gk\left[\rho^4\right.\right.\\
&\quad \left.\left. + 2\rho^2\left(1-2\sigma^2\right)\chi^2 + \rho\left(1-2\sigma^2-3\sigma^4\right)\chi^3 + \left(1-5\sigma^2+3\sigma^4\right)\chi^4 + \rho^3(\ \chi + 2\sigma^2\chi)\right]\right\}\sin\left(3kx - 3\omega_0\tau\right).
\end{aligned}
\qquad [26]
$$

Following Zhao & Liu (1), the parameter $\beta_2$ is used to avoid the secular term on the right-hand side. In other words, $\beta_2$ is adopted to eliminate the $\sin\left(kx - \omega_0\tau\right)$ term in the forcing term. Therefore, the following expression should be forced:

$$
\beta_2 = \frac{k^2A^2\left[\rho^4 + 2\rho^3\left(1+\sigma^2\right)\chi - 3\rho^2\left(-1+\sigma^2\right)\chi^2 + 2\rho\left(-1+\sigma^2\right)^2\chi^3 + \left(1-5\sigma^2+7\sigma^4\right)\chi^4\right]}{8\omega_0\sigma^4\chi^2(\rho+\chi)}. \qquad [27]
$$

Then, $\eta_3$ can be obtained once $\phi_3$ is known.

$$
\begin{aligned}
\eta_3 = -\frac{1}{2g}\{2[&\frac{\partial\phi_1}{\partial x}\frac{\partial\phi_2}{\partial x} + \eta_2\left(\frac{\partial^2\phi_1}{\partial z\partial\tau} + \Omega\frac{\partial\phi_1}{\partial x} - \Omega\frac{\partial\psi_1}{\partial z} + U_0\frac{\partial^2\phi_1}{\partial x\partial z}\right) - \Omega\psi_3 + \beta_2\frac{\partial\phi_1}{\partial\tau} + \frac{\partial\phi_3}{\partial\tau}\\
&+\frac{\partial\phi_1}{\partial z}\frac{\partial\phi_2}{\partial z} + U_0\frac{\partial\phi_3}{\partial x}\Big]\\
&+ 2\eta_1\left(\frac{\partial^2\phi_1}{\partial x\partial z}\frac{\partial\phi_1}{\partial x} + \frac{\partial^2\phi_1}{\partial z^2}\frac{\partial\phi_1}{\partial z} + \Omega\frac{\partial\phi_2}{\partial x} - \Omega\frac{\partial\psi_2}{\partial z} + \frac{\partial^2\phi_2}{\partial z\partial\tau} + U_0\frac{\partial^2\phi_2}{\partial x\partial z}\right)\\
&+\eta_1^2\left(2\Omega\frac{\partial^2\phi_1}{\partial x\partial z} - \Omega\frac{\partial^2\psi_1}{\partial z^2} + \frac{\partial^3\phi_1}{\partial z^2\partial\tau} + U_0\frac{\partial^3\phi_1}{\partial x\partial z^2}\right)\Big\}.
\end{aligned}
\qquad [28]
$$

Thus, the third order solutions can be obtained:

$$
\phi_3 = A_{33}\cosh 3k(z+h)\sin\left(3kx - 3\omega_0\tau\right), \qquad [29]
$$

$$
\psi_3 = A_{33}\sinh 3k(z+h)\cos\left(3kx - 3\omega_0\tau\right), \qquad [30]
$$

$$
\eta_3 = B_{31}\cos\left(kx - \omega_0\tau\right) + B_{33}\cos\left(3kx - 3\omega_0\tau\right), \qquad [31]
$$

where

$$
A_{33} = \frac{A^3k\left(1-\sigma^2\right)^{3/2}}{64\sigma^7\chi^3}\left[\rho^4 + 2\rho^3\left(1+\sigma^2\right)\chi + \rho^2\left(3-7\sigma^2\right)\chi^2 - 2\rho\left(-1+4\sigma^2+\sigma^4\right)\chi^3 + \left(1-9\sigma^2+15\sigma^4\right)\chi^4\right], \qquad [32]
$$

$$
B_{31} = \frac{A^3k^2\left[\rho^3 + \rho\left(1+\sigma^2-2\sigma^4\right)\chi^2 + \sigma^2\left(3-7\sigma^2\right)\chi^3 + \rho^2\chi\left(1+4\sigma^2\right)\right]}{8\sigma^4\chi^2(\rho+\chi)}, \qquad [33]
$$

$$
\begin{aligned}
B_{33} = &\frac{A^3k^2}{64\sigma^6\chi^4}\left[\rho^4\left(3+\sigma^2\right) + 2\rho^3\left(3+8\sigma^2+\sigma^4\right)\chi + \rho^2\left(9-2\sigma^2+9\sigma^4\right)\chi^2\right.\\
&\left. -2\rho\left(-3+3\sigma^2+11\sigma^4+\sigma^6\right)\chi^3 - \left(-3+18\sigma^2-20\sigma^4+\sigma^6\right)\chi^4\right].
\end{aligned}
\qquad [34]
$$

**D. Four-order equations and solutions.** For the forth-order problem, the combined and the dynamic free surface boundary conditions at $z = 0$ are given as follows:

$$\begin{aligned}
&\frac{\partial^2\phi_4}{\partial\tau^2} + g\frac{\partial\phi_4}{\partial z} + U_0^2\frac{\partial^2\phi_4}{\partial x^2} + 2U_0\frac{\partial^2\phi_4}{\partial x\partial\tau} - \Omega\left(\frac{\partial\psi_4}{\partial\tau} + U_0\frac{\partial\psi_4}{\partial x}\right) \\
&\quad= -\frac{k^2A^4}{192\sigma^8\chi^3(\rho+\chi)}\left\{3\rho^7\left(-1+\sigma^2\right) + 6\rho^6\left(-2+5\sigma^2+\sigma^4\right)\chi + 9\rho^5(-3\right. \\
&\qquad +9\sigma^2+2\sigma^4\big)\chi^2 + 3\rho^4\left(-13+16\sigma^2+45\sigma^4-16\sigma^6\right)\chi^3 - 3\rho^3(13 \\
&\qquad +11\sigma^2-66\sigma^4+10\sigma^6\big)\chi^4 + 3\rho^2\big(-9-39\sigma^2+193\sigma^4-135\sigma^6 \\
&\qquad +14\sigma^8\big)\chi^5 + \rho\left(-12-87\sigma^2+522\sigma^4-454\sigma^6+55\sigma^8\right)\chi^6 + (-1+\sigma)(1 \\
&\qquad \left.+\sigma)\left(3+36\sigma^2-234\sigma^4+193\sigma^6\right)\chi^7\right\}\sin\left(2kx-2\omega_0\tau\right) - \frac{k^2A^4}{96\sigma^8\chi^3}\left\{3\rho^6(5\right. \\
&\qquad +\sigma^2\big) + 3\rho^5\left(15+33\sigma^2+2\sigma^4\right)\chi + 6\rho^4\left(15-12\sigma^2+11\sigma^4\right)\chi^2 \\
&\qquad -3\rho^3\left(-35+98\sigma^2+169\sigma^4+28\sigma^6\right)\chi^3 + 3\rho^2\big(30-219\sigma^2+187\sigma^4 \\
&\qquad +24\sigma^6\big)\chi^4 + 3\rho\left(15-162\sigma^2+252\sigma^4+207\sigma^6+26\sigma^8\right)\chi^5 + (15 \\
&\qquad \left.-267\sigma^2+1134\sigma^4-1159\sigma^6-275\sigma^8\big)\chi^6\right\}\sin\left(4kx-4\omega_0\tau\right),
\end{aligned} \tag{35}$$

and

$$\begin{aligned}
\eta_4 =& -\frac{A^4k^3\left(-1+\sigma^2\right)}{64\omega_0\rho\sigma^7\chi^3(\rho+\chi)}\left\{\omega_0\left(-1+\sigma^2\right)(\rho+\chi)\left(\rho^2+\rho\chi+\left(1-3\sigma^2\right)\chi^2\right)^2\right. \\
&+2\sigma^2\chi^2\left(\rho^4+2\rho^3\left(1+\sigma^2\right)\chi-3\rho^2\left(-1+\sigma^2\right)\chi^2+2\rho\left(-1+\sigma^2\right)^2\chi^3\right. \\
&\left.\left.+\left(1-5\sigma^2+7\sigma^4\right)\chi^4\right)\right\} + \frac{\sigma}{\rho\chi}\left(k\Omega\psi_4 - k\frac{\partial\phi_4}{\partial\tau} + (-\omega_0+\chi)\frac{\partial\phi_4}{\partial x}\right) \\
&+\frac{A^4k^3}{384\rho\sigma^7\chi^4(\rho+\chi)}\left\{12\rho^6\left(1+\sigma^2\right) + 3\rho^5\left(9+32\sigma^2+15\sigma^4\right)\chi\right. \\
&+3\rho^4\left(15+38\sigma^2-3\sigma^4+14\sigma^6\right)\chi^2 \\
&+3\rho^3\left(13+59\sigma^2-149\sigma^4-19\sigma^6\right)\chi^3 \\
&+\rho^2\left(27+135\sigma^2-645\sigma^4+245\sigma^6-42\sigma^8\right)\chi^4 \\
&+\rho\left(9+117\sigma^2-690\sigma^4+675\sigma^6-55\sigma^8\right)\chi^5 \\
&\left.+\left(3+33\sigma^2-270\sigma^4+427\sigma^6-193\sigma^8\right)\chi^6\right\}\cos\left(2kx-2\omega_0\tau\right) \\
&+\frac{A^4k^3}{384\rho\sigma^7\chi^3}\left\{3\rho^4\left(-5+50\sigma^2+11\sigma^4\right) + 6\rho^3\big(-5+47\sigma^2+41\sigma^4\right. \\
&+13\sigma^6\big)\chi - 3\rho^2\left(15-183\sigma^2+233\sigma^4+47\sigma^6\right)\chi^2 - 2\rho\big(15-204\sigma^2 \\
&+402\sigma^4+164\sigma^6+39\sigma^8\big)\chi^3 + \big(-15+267\sigma^2-1134\sigma^4+1159\sigma^6 \\
&\left.+275\sigma^8\big)\chi^4\right\}\cos\left(4kx-4\omega_0\tau\right).
\end{aligned} \tag{36}$$

The corresponding forth-order solutions can be expressed as:

$$\phi_4 = C_4\tau + A_{42}\cosh 2k(z+h)\sin\left(2kx-2\omega_0\tau\right) + A_{44}\cosh 4k(z+h)\sin\left(4kx-4\omega_0\tau\right), \tag{37}$$

$$\psi_4 = A_{42}\sinh 2k(z+h)\cos\left(2kx-2\omega_0\tau\right) + A_{44}\sinh 4k(z+h)\cos\left(4kx-4\omega_0\tau\right), \tag{38}$$

$$\begin{aligned}
\eta_4 =& B_{42}\cos\left(2kx-2\omega_0\tau\right) + B_{44}\cos\left(4kx-4\omega_0\tau\right) - \frac{C_4k\sigma}{\rho\chi} \\
&-\frac{A^4k^3\left(-1+\sigma^2\right)}{64\omega_0\rho\sigma^7\chi^3(\rho+\chi)}\left\{\omega_0\left(-1+\sigma^2\right)(\rho+\chi)\left[\rho^2+\rho\chi+\left(1-3\sigma^2\right)\chi^2\right]^2 + 2\sigma^2\chi^2\left[\rho^4+2\rho^3\left(1+\sigma^2\right)\chi\right.\right. \\
&\left.\left.-3\rho^2\left(-1+\sigma^2\right)\chi^2 + 2\rho\left(-1+\sigma^2\right)^2\chi^3 + \left(1-5\sigma^2+7\sigma^4\right)\chi^4\right]\right\}.
\end{aligned} \tag{39}$$

where

$$A_{42} = \frac{k^2 A^4 \left(1-\sigma^2\right)}{768\sigma^{10}\chi^5(\rho+\chi)} \left\{3\rho^7\left(-1+\sigma^2\right)+6\rho^6\left(-2+5\sigma^2+\sigma^4\right)\chi \right. \\ +9\rho^5\left(-3+9\sigma^2+2\sigma^4\right)\chi^2+3\rho^4\left(-13+16\sigma^2+45\sigma^4-16\sigma^6\right)\chi^3 \\ -3\rho^3\left(13+11\sigma^2-66\sigma^4+10\sigma^6\right)\chi^4 \\ +3\rho^2\left(-9-39\sigma^2+193\sigma^4-135\sigma^6+14\sigma^8\right)\chi^5 \\ +\rho\left(-12-87\sigma^2+522\sigma^4-454\sigma^6+55\sigma^8\right)\chi^6 \\ \left. +(-1+\sigma)(1+\sigma)\left(3+36\sigma^2-234\sigma^4+193\sigma^6\right)\chi^7\right\}, \quad [40]$$

$$A_{44} = \frac{k^2 A^4 \left(-1+\sigma^2\right)^2}{1536\sigma^{10}\left(5+\sigma^2\right)\chi^5} \left\{3\rho^6\left(5+\sigma^2\right)+3\rho^5\left(15+33\sigma^2+2\sigma^4\right)\chi \right. \\ +6\rho^4\left(15-12\sigma^2+11\sigma^4\right)\chi^2-3\rho^3\left(-35+98\sigma^2+169\sigma^4+28\sigma^6\right)\chi^3 \\ +3\rho^2\left(30-219\sigma^2+187\sigma^4+24\sigma^6\right)\chi^4 \\ +3\rho\left(15-162\sigma^2+252\sigma^4+207\sigma^6+26\sigma^8\right)\chi^5 \\ \left. +\left(15-267\sigma^2+1134\sigma^4-1159\sigma^6-275\sigma^8\right)\chi^6\right\}, \quad [41]$$

$$B_{42} = \frac{k^3 A^4}{384\sigma^9\chi^6(\rho+\chi)} \left\{3\rho^7\left(-1+\sigma^2\right)+3\rho^6\left(-4+9\sigma^2+3\sigma^4\right)\chi+3\rho^5\left(-9+27\sigma^2 \right.\right. \\ \left.+20\sigma^4+2\sigma^6\right)\chi^2+3\rho^4\left(-13+16\sigma^2+104\sigma^4+5\sigma^6\right)\chi^3-3\rho^3(13 \\ +9\sigma^2-120\sigma^4-32\sigma^6+2\sigma^8)\chi^4-3\rho^2\left(9+39\sigma^2-241\sigma^4+218\sigma^6\right. \\ \left.+15\sigma^8\right)\chi^5-\rho\left(12+87\sigma^2-540\sigma^4+520\sigma^6+105\sigma^8\right)\chi^6+\left(-3-36\sigma^2\right. \\ \left.\left.+300\sigma^4-595\sigma^6+414\sigma^8\right)\chi^7\right\}, \quad [42]$$

$$B_{44} = \frac{k^3 A^4}{384\sigma^9\left(5+\sigma^2\right)\chi^6} \left\{3\rho^6\left(5+6\sigma^2+\sigma^4\right)+3\rho^5\left(15+73\sigma^2+45\sigma^4+3\sigma^6\right)\chi \right. \\ +3\rho^4\left(30+81\sigma^2+178\sigma^4+65\sigma^6+2\sigma^8\right)\chi^2 \\ +3\rho^3\left(35+62\sigma^2-112\sigma^4-6\sigma^6+5\sigma^8\right)\chi^3 \\ -3\rho^2\left(-30+64\sigma^2+27\sigma^4+228\sigma^6+73\sigma^8+2\sigma^{10}\right)\chi^4 \\ +3\rho\left(15-72\sigma^2-75\sigma^4+193\sigma^6+72\sigma^8+3\sigma^{10}\right)\chi^5 \\ \left. +\left(15-177\sigma^2+(-2+\sigma)\sigma^4(2+\sigma)\left(-3+2\sigma^2\right)\left(41+9\sigma^2\right)\right)\chi^6\right\}, \quad [43]$$

$$C_4 = -\frac{A^4 k^2\left(-1+\sigma^2\right)}{64\omega_0\sigma^8\chi^2(\rho+\chi)} \left\{\omega_0\left(-1+\sigma^2\right)(\rho+\chi)\left[\rho^2+\rho\chi+\left(1-3\sigma^2\right)\chi^2\right]^2+2\sigma^2\chi^2\left[\rho^4 \right.\right. \\ \left.\left. +2\rho^3\left(1+\sigma^2\right)\chi-3\rho^2\left(-1+\sigma^2\right)\chi^2+2\rho\left(-1+\sigma^2\right)^2\chi^3+\left(1-5\sigma^2+7\sigma^4\right)\chi^4\right]\right\}. \quad [44]$$

**E. Fifth-order equations and solutions.** For the fifth-order problem, the combined condition at $z=0$ gives:

$$\frac{\partial^2\phi_5}{\partial\tau^2}+g\frac{\partial\phi_5}{\partial z}+U_0^2\frac{\partial^2\phi_5}{\partial x^2}+2U_0\frac{\partial^2\phi_5}{\partial x\partial\tau}-\Omega\left(\frac{\partial\psi_5}{\partial\tau}+U_0\frac{\partial\psi_5}{\partial x}\right) \\ =\zeta_1\sin\left(kx-\omega_0\tau\right)+\zeta_3\sin\left(3kx-3\omega_0\tau\right)+\zeta_5\sin\left(5kx-5\omega_0\tau\right), \quad [45]$$

where

$$\zeta_1 = -\frac{k^3A^5}{512\sigma^{11}\chi^5(\rho+\chi)^2}\left[\rho^{10}\left(1+3\sigma^2\right)+6\rho^9\left(1+9\sigma^2+2\sigma^4\right)\chi+\rho^8\left(19+203\sigma^2+146\sigma^4+12\sigma^6\right)\chi^2+8\rho^7\left(5+42\sigma^2\right.\right. \\ \left.+56\sigma^4+9\sigma^6\right)\chi^3+\rho^6\left(61+197\sigma^2+733\sigma^4+145\sigma^6-24\sigma^8\right)\chi^4-2\rho^5\left(-35+153\sigma^2-695\sigma^4\right. \\ \left.+237\sigma^6+28\sigma^8\right)\chi^5+\rho^4\left(61-849\sigma^2+2663\sigma^4-1749\sigma^6-98\sigma^8+12\sigma^{10}\right)\chi^6-4\rho^3(-1+\sigma)(1 \\ +\sigma)\left(10-238\sigma^2+669\sigma^4-198\sigma^6+23\sigma^8\right)\chi^7+\rho^2\left(19-701\sigma^2+3285\sigma^4-4755\sigma^6+2789\sigma^8\right. \\ \left.-729\sigma^{10}\right)\chi^8-6\rho\left(-1+49\sigma^2-285\sigma^4+545\sigma^6-445\sigma^8+189\sigma^{10}\right)\chi^9+\left(1-63\sigma^2+457\sigma^4\right. \\ \left.\left.-1161\sigma^6+1311\sigma^8-709\sigma^{10}\right)\chi^{10}\right]+\frac{A\omega_0\chi(\rho+\chi)\beta_4}{k\sigma}, \quad [46]$$

 Haiqi Fang, Philip L.-F. Liu, Lian Tang, Pengzhi Lin

$$
\begin{aligned}
\zeta_3 =& \frac{3k^3A^5}{512\sigma^{11}\left(5+\sigma^2\right)\chi^5(\rho+\chi)}\left[-3\rho^9\left(-5+4\sigma^2+\sigma^4\right)-\rho^8\left(5+3\sigma^2\right)\left(-15+33\sigma^2+4\sigma^4\right)\chi\right.\\
&-2\rho^7\left(-105+249\sigma^2+319\sigma^4+99\sigma^6+6\sigma^8\right)\chi^2-2\rho^6\left(-195+351\sigma^2+773\sigma^4-27\sigma^6+2\sigma^8\right)\chi^3\\
&+\rho^5\left(525+\sigma^2\left(-2+\sigma^2\right)\left(240+654\sigma^2+591\sigma^4+16\sigma^6\right)\right)\chi^4\\
&+\rho^4\left(525+420\sigma^2-4944\sigma^4+4716\sigma^6-397\sigma^8-400\sigma^{10}\right)\chi^5\\
&+2\rho^3\left(195+504\sigma^2-4549\sigma^4+2687\sigma^6+1648\sigma^8-43\sigma^{10}-2\sigma^{12}\right)\chi^6\\
&+2\rho^2\left(105+516\sigma^2-6185\sigma^4+10357\sigma^6-3680\sigma^8-485\sigma^{10}+148\sigma^{12}\right)\chi^7\\
&+\rho\left(75+510\sigma^2-7930\sigma^4+19440\sigma^6-9366\sigma^8-2082\sigma^{10}+73\sigma^{12}\right)\chi^8\\
&\left.+3(-1+\sigma)(1+\sigma)\left(-5-51\sigma^2+919\sigma^4-2833\sigma^6+1509\sigma^8+431\sigma^{10}\right)\chi^9\right],
\end{aligned}
\tag{47}
$$

$$
\begin{aligned}
\zeta_5 =& -\frac{5k^3A^5}{1536\sigma^{11}\left(5+\sigma^2\right)\chi^5}\left[3\rho^8\left(5+\sigma^2\right)\left(5+3\sigma^2\right)+12\rho^7\left(25+95\sigma^2+41\sigma^4+3\sigma^6\right)\chi\right.\\
&+6\rho^6\left(125+175\sigma^2+429\sigma^4+129\sigma^6+6\sigma^8\right)\chi^2-24\rho^5\left(-50+35\sigma^2+350\sigma^4+154\sigma^6+\sigma^8\right)\chi^3\\
&-3\rho^4\left(-475+2320\sigma^2+1844\sigma^4+3512\sigma^6+1871\sigma^8+48\sigma^{10}\right)\chi^4\\
&+12\rho^3\left(100-895\sigma^2+69\sigma^4+2701\sigma^6+1677\sigma^8+216\sigma^{10}\right)\chi^5\\
&+2\rho^2\left(375-5550\sigma^2+14289\sigma^4+2737\sigma^6-3022\sigma^8+813\sigma^{10}+54\sigma^{12}\right)\chi^6\\
&-4\rho\left(-75+1515\sigma^2-6273\sigma^4+1066\sigma^6+11321\sigma^8+5781\sigma^{10}+621\sigma^{12}\right)\chi^7\\
&\left.+5\left(15-438\sigma^2+3330\sigma^4-7808\sigma^6+1694\sigma^8+6450\sigma^{10}+1125\sigma^{12}\right)\chi^8\right].
\end{aligned}
\tag{48}
$$

To eliminate the $\sin\left(kx-\omega_0\tau\right)$ term on the right-hand side of the combined boundary condition (in $\zeta_1$, equation (46)), $\beta_4$ is set as follows:

$$
\begin{aligned}
\beta_4 =& \frac{k^4A^4}{512\omega_0\sigma^{10}\chi^6(\rho+\chi)^3}\left\{\rho^{10}\left(1+3\sigma^2\right)+6\rho^9\left(1+9\sigma^2+2\sigma^4\right)\chi+\rho^8\left(19+203\sigma^2+146\sigma^4+12\sigma^6\right)\chi^2+8\rho^7\left(5+42\sigma^2\right.\right.\\
&\left.+56\sigma^4+9\sigma^6\right)\chi^3+\rho^6\left(61+197\sigma^2+733\sigma^4+145\sigma^6-24\sigma^8\right)\chi^4-2\rho^5\left(-35+153\sigma^2-695\sigma^4\right.\\
&\left.+237\sigma^6+28\sigma^8\right)\chi^5+\rho^4\left(61-849\sigma^2+2663\sigma^4-1749\sigma^6-98\sigma^8+12\sigma^{10}\right)\chi^6-4\rho^3(-1+\sigma)(1\\
&+\sigma)\left(10-238\sigma^2+669\sigma^4-198\sigma^6+23\sigma^8\right)\chi^7+\rho^2\left(19-701\sigma^2+3285\sigma^4-4755\sigma^6+2789\sigma^8\right.\\
&\left.-729\sigma^{10}\right)\chi^8-6\rho\left(-1+49\sigma^2-285\sigma^4+545\sigma^6-445\sigma^8+189\sigma^{10}\right)\chi^9+\left(1-63\sigma^2+457\sigma^4\right.\\
&\left.\left.-1161\sigma^6+1311\sigma^8-709\sigma^{10}\right)\chi^{10}\right\}.
\end{aligned}
\tag{49}
$$

Thus the secular behavior in the fifth-order problem for $\phi_5$ can be eliminated, and the value of $\beta$ in equations (5) and (6) can be determined.

The dynamic condition gives

$$
\eta_5=\xi_1\cos\left(kx-\omega_0\tau\right)+\xi_3\cos\left(3kx-3\omega_0\tau\right)+\xi_5\cos\left(5kx-5\omega_0\tau\right)+\frac{\sigma}{\rho\chi}\left[k\Omega\psi_5-k\frac{\partial\phi_5}{\partial\tau}+\left(-\omega_0+\chi\right)\frac{\partial\phi_5}{\partial x}\right], \tag{50}
$$

where

$$
\begin{aligned}
\xi_1 =& \frac{A^5k^4}{1536\sigma^{10}\chi^6(\rho+\chi)^3}\left\{\rho^9\left(3+9\sigma^2\right)+3\rho^8\left(5+57\sigma^2+14\sigma^4\right)\chi+3\rho^7\left(13+203\sigma^2+188\sigma^4+16\sigma^6\right)\chi^2\right.\\
&+3\rho^6\left(22+336\sigma^2+529\sigma^4+109\sigma^6\right)\chi^3+\rho^5\left(78+969\sigma^2+1425\sigma^4+1166\sigma^6-202\sigma^8\right)\chi^4\\
&+\rho^4\left(66+522\sigma^2+48\sigma^4+1647\sigma^6-1567\sigma^8\right)\chi^5\\
&+\rho^3\left(39+126\sigma^2-1413\sigma^4+4383\sigma^6-5773\sigma^8+426\sigma^{10}\right)\chi^6\\
&+\rho^2\left(15+48\sigma^2-2220\sigma^4+7772\sigma^6-9400\sigma^8+2901\sigma^{10}\right)\chi^7\\
&+\rho\left(3+57\sigma^2-1512\sigma^4+5883\sigma^6-8179\sigma^8+5124\sigma^{10}\right)\chi^8\\
&\left.+\sigma^2\left(45-681\sigma^2+2838\sigma^4-4627\sigma^6+3249\sigma^8\right)\chi^9\right\},
\end{aligned}
\tag{51}
$$

$$
\begin{aligned}
\xi_3 =& \frac{3k^4A^5}{512\rho\sigma^{10}\left(5+\sigma^2\right)\chi^6(\rho+\chi)}\left\{\rho^8\left(5+\sigma^2\right)\left(1+3\sigma^2\right)+4\rho^7\left(5+\sigma^2\right)\left(1+4\sigma^2+6\sigma^4\right)\chi\right.\\
&+2\rho^6\left(25+55\sigma^2+208\sigma^4+185\sigma^6+15\sigma^8\right)\chi^2\\
&+2\rho^5\left(40+48\sigma^2+135\sigma^4+18\sigma^6+53\sigma^8-6\sigma^{10}\right)\chi^3\\
&+\rho^4\left(95+84\sigma^2-922\sigma^4+514\sigma^6-609\sigma^8-178\sigma^{10}\right)\chi^4\\
&+2\rho^3\left(40+78\sigma^2-1249\sigma^4+967\sigma^6+245\sigma^8-43\sigma^{10}+6\sigma^{12}\right)\chi^5\\
&+2\rho^2\left(25+100\sigma^2-1750\sigma^4+3330\sigma^6-1126\sigma^8-76\sigma^{10}+61\sigma^{12}\right)\chi^6\\
&+2\rho\left(10+72\sigma^2-1263\sigma^4+3411\sigma^6-1912\sigma^8-269\sigma^{10}+43\sigma^{12}\right)\chi^7\\
&\left.+(-1+\sigma)(1+\sigma)\left(-5-51\sigma^2+919\sigma^4-2833\sigma^6+1509\sigma^8+431\sigma^{10}\right)\chi^8\right\},
\end{aligned}
\tag{52}
$$

$$
\begin{aligned}
\xi_5 =& \frac{5A^5k^4}{1536\rho\sigma^{10}\left(5+\sigma^2\right)\chi^5}\left\{3\rho^6\left(5+\sigma^2\right)\left(-1+15\sigma^2+10\sigma^4\right)+3\rho^5\left(-15+204\sigma^2+463\sigma^4+320\sigma^6+20\sigma^8\right)\chi\right.\\
&+3\rho^4\left(-30+480\sigma^2+21\sigma^4+60\sigma^6+205\sigma^8\right)\chi^2-\rho^3\left(105-1851\sigma^2+1731\sigma^4+5257\sigma^6\right.\\
&\left.+3780\sigma^8+530\sigma^{10}\right)\chi^3+\rho^2\left(-90+1899\sigma^2-6231\sigma^4+330\sigma^6+2885\sigma^8+55\sigma^{10}\right)\chi^4+\rho(-45\\
&+1071\sigma^2-5034\sigma^4+2541\sigma^6+7133\sigma^8+3940\sigma^{10}+450\sigma^{12}\big)\chi^5-\left(15-438\sigma^2+3330\sigma^4\right.\\
&\left.\left.-7808\sigma^6+1694\sigma^8+6450\sigma^{10}+1125\sigma^{12}\right)\chi^6\right\}.
\end{aligned}
\tag{53}
$$

Then the fifth order solutions are obtained,

$$
\begin{aligned}
\phi_5 =& A_{53}\cosh 3k(z+h)\sin\left(3kx-3\omega_0\tau\right)+A_{55}\cosh 5k(z+h)\sin\left(5kx-5\omega_0\tau\right),\\
\psi_5 =& A_{53}\sinh 3k(z+h)\cos\left(3kx-3\omega_0\tau\right)+A_{55}\sinh 5k(z+h)\cos\left(5kx-5\omega_0\tau\right),\\
\eta_5 =& B_{51}\cos\left(kx-\omega_0\tau\right)+B_{53}\cos\left(3kx-3\omega_0\tau\right)+B_{55}\cos\left(5kx-5\omega_0\tau\right),
\end{aligned}
\tag{54}
$$

where

$$
\begin{aligned}
A_{53} =& \frac{k^3A^5\left(1-\sigma^2\right)^{3/2}}{4096\sigma^{13}\left(5+\sigma^2\right)\chi^7(\rho+\chi)}\left\{3\rho^9\left(-5+4\sigma^2+\sigma^4\right)+\rho^8\left(5+3\sigma^2\right)\left(-15+33\sigma^2+4\sigma^4\right)\chi\right.\\
&+2\rho^7\left(-105+249\sigma^2+319\sigma^4+99\sigma^6+6\sigma^8\right)\chi^2+2\rho^6\left(-195+351\sigma^2+773\sigma^4-27\sigma^6+2\sigma^8\right)\chi^3\\
&+\rho^5\left(-525-\sigma^2\left(-2+\sigma^2\right)\left(240+654\sigma^2+591\sigma^4+16\sigma^6\right)\right)\chi^4\\
&+\rho^4\left(-525-420\sigma^2+4944\sigma^4-4716\sigma^6+397\sigma^8+400\sigma^{10}\right)\chi^5\\
&+2\rho^3\left(-195-504\sigma^2+4549\sigma^4-2687\sigma^6-1648\sigma^8+43\sigma^{10}+2\sigma^{12}\right)\chi^6\\
&-2\rho^2\left(105+516\sigma^2-6185\sigma^4+10357\sigma^6-3680\sigma^8-485\sigma^{10}+148\sigma^{12}\right)\chi^7\\
&+\rho\left(-75-510\sigma^2+7930\sigma^4-19440\sigma^6+9366\sigma^8+2082\sigma^{10}-73\sigma^{12}\right)\chi^8\\
&\left.-3(-1+\sigma)(1+\sigma)\left(-5-51\sigma^2+919\sigma^4-2833\sigma^6+1509\sigma^8+431\sigma^{10}\right)\chi^9\right\},
\end{aligned}
\tag{55}
$$

$$
\begin{aligned}
A_{55} =& \frac{k^3A^5\left(1-\sigma^2\right)^{5/2}}{12288\sigma^{13}\left(5+\sigma^2\right)\left(5+3\sigma^2\right)\chi^7}\left\{3\rho^8\left(5+\sigma^2\right)\left(5+3\sigma^2\right)+12\rho^7\left(25+95\sigma^2+41\sigma^4+3\sigma^6\right)\chi+6\rho^6\left(125+175\sigma^2\right.\right.\\
&\left.+429\sigma^4+129\sigma^6+6\sigma^8\right)\chi^2-24\rho^5\left(-50+35\sigma^2+350\sigma^4+154\sigma^6+\sigma^8\right)\chi^3-3\rho^4\left(-475+2320\sigma^2\right.\\
&\left.+1844\sigma^4+3512\sigma^6+1871\sigma^8+48\sigma^{10}\right)\chi^4+12\rho^3\left(100-895\sigma^2+69\sigma^4+2701\sigma^6+1677\sigma^8\right.\\
&\left.+216\sigma^{10}\right)\chi^5+2\rho^2\left(375-5550\sigma^2+14289\sigma^4+2737\sigma^6-3022\sigma^8+813\sigma^{10}+54\sigma^{12}\right)\chi^6\\
&-4\rho\left(-75+1515\sigma^2-6273\sigma^4+1066\sigma^6+11321\sigma^8+5781\sigma^{10}+621\sigma^{12}\right)\chi^7+5\left(15-438\sigma^2\right.\\
&\left.\left.+3330\sigma^4-7808\sigma^6+1694\sigma^8+6450\sigma^{10}+1125\sigma^{12}\right)\chi^8\right\},
\end{aligned}
\tag{56}
$$

$$
\begin{aligned}
B_{51} =& \frac{k^4A^5}{1536\sigma^{10}\chi^6(\rho+\chi)^3}\left\{\rho^9\left(3+9\sigma^2\right)+3\rho^8\left(5+57\sigma^2+14\sigma^4\right)\chi+3\rho^7\left(13+203\sigma^2+188\sigma^4+16\sigma^6\right)\chi^2+3\rho^6(22\right.\\
&+336\sigma^2+529\sigma^4+109\sigma^6\big)\chi^3+\rho^5\left(78+969\sigma^2+1425\sigma^4+1166\sigma^6-202\sigma^8\right)\chi^4+\rho^4\left(66+522\sigma^2\right.\\
&\left.+48\sigma^4+1647\sigma^6-1567\sigma^8\right)\chi^5+\rho^3\left(39+126\sigma^2-1413\sigma^4+4383\sigma^6-5773\sigma^8+426\sigma^{10}\right)\chi^6\\
&+\rho^2\left(15+48\sigma^2-2220\sigma^4+7772\sigma^6-9400\sigma^8+2901\sigma^{10}\right)\chi^7+\rho\left(3+57\sigma^2-1512\sigma^4+5883\sigma^6\right.\\
&\left.\left.-8179\sigma^8+5124\sigma^{10}\right)\chi^8+\sigma^2\left(45-681\sigma^2+2838\sigma^4-4627\sigma^6+3249\sigma^8\right)\chi^9\right\},
\end{aligned}
\tag{57}
$$

$$
\begin{aligned}
B_{53} =& \frac{k^4 A^5}{4096\sigma^{12}\left(5+\sigma^2\right)\chi^8(\rho+\chi)} \left\{3\rho^9\left(-15+7\sigma^2+7\sigma^4+\sigma^6\right)+\rho^8\left(-225+165\sigma^2+573\sigma^4+179\sigma^6+12\sigma^8\right)\chi\right.\\
&+6\rho^7\left(-105+134\sigma^2+626\sigma^4+376\sigma^6+55\sigma^8+2\sigma^{10}\right)\chi^2+2\rho^6\left(-585+258\sigma^2+5670\sigma^4\right.\\
&\left.+4876\sigma^6+1059\sigma^8+50\sigma^{10}\right)\chi^3+\rho^5\left(-1575-1005\sigma^2+11940\sigma^4+25004\sigma^6+7299\sigma^8\right.\\
&\left.+145\sigma^{10}-16\sigma^{12}\right)\chi^4+\rho^4\left(-1575-4065\sigma^2+20556\sigma^4+5820\sigma^6+1563\sigma^8-331\sigma^{10}\right.\\
&\left.-16\sigma^{12}\right)\chi^5+2\rho^3\left(-585-2667\sigma^2+12471\sigma^4+5200\sigma^6-20327\sigma^8-7239\sigma^{10}-487\sigma^{12}\right.\\
&\left.+2\sigma^{14}\right)\chi^6+2\rho^2\left(-315-2253\sigma^2+15879\sigma^4-18470\sigma^6+2395\sigma^8-2169\sigma^{10}-647\sigma^{12}\right.\\
&\left.+12\sigma^{14}\right)\chi^7+\rho\left(-225-2085\sigma^2+19824\sigma^4-35430\sigma^6+2786\sigma^8+20444\sigma^{10}+5975\sigma^{12}\right.\\
&\left.\left.+487\sigma^{14}\right)\chi^8+\left(-45-549\sigma^2+7968\sigma^4-28042\sigma^6+36030\sigma^8-13524\sigma^{10}-3369\sigma^{12}+187\sigma^{14}\right)\chi^9\right\},
\end{aligned}
\quad [58]
$$

$$
\begin{aligned}
B_{55} =& \frac{k^4 A^5}{12288\sigma^{12}\left(5+\sigma^2\right)\left(5+3\sigma^2\right)\chi^8} \left\{3\rho^8\left(5+\sigma^2\right)\left(5+3\sigma^2\right)\left(5+10\sigma^2+\sigma^4\right)+12\rho^7\left(125+975\sigma^2+1530\sigma^4+670\sigma^6\right.\right.\\
&\left.+89\sigma^8+3\sigma^{10}\right)\chi+6\rho^6\left(1+\sigma^2\right)\left(625+3500\sigma^2+9320\sigma^4+3630\sigma^6+327\sigma^8+6\sigma^{10}\right)\chi^2\\
&+24\rho^5\left(250+1450\sigma^2+2225\sigma^4+3770\sigma^6+3194\sigma^8+820\sigma^{10}+35\sigma^{12}\right)\chi^3-3\rho^4(-2375\\
&-6150\sigma^2-5455\sigma^4+39960\sigma^6+43239\sigma^8+9950\sigma^{10}+143\sigma^{12}+48\sigma^{14})\chi^4-12\rho^3(-500\\
&+225\sigma^2+5755\sigma^4+3610\sigma^6+7090\sigma^8+7181\sigma^{10}+1719\sigma^{12}+72\sigma^{14})\chi^5+2\rho^2\left(1875-10500\sigma^2\right.\\
&\left.-5280\sigma^4-23335\sigma^6+55165\sigma^8+74586\sigma^{10}+18906\sigma^{12}+657\sigma^{14}+54\sigma^{16}\right)\chi^6+4\rho(375\\
&-3825\sigma^2+2040\sigma^4+20485\sigma^6-5214\sigma^8-16627\sigma^{10}-2940\sigma^{12}+495\sigma^{14}+27\sigma^{16})\chi^7+(375\\
&\left.-7200\sigma^2+35025\sigma^4-50930\sigma^6+4368\sigma^8+24734\sigma^{10}-8721\sigma^{12}-3036\sigma^{14}+9\sigma^{16})\chi^8\right\}.
\end{aligned}
\quad [59]
$$

## 2. Software resources

The derivation of the fifth-order solutions of Stokes waves in a linear shear current is compeleted with the aid of Mathematica®. A software resource can be downloaded from the link: https://github.com/474278604/Haiqi-Fang.git, in which one can use the files to derive the fifth-order solutions starting from the first-order problems and can also extend to higher-order solutions based on the present framework. In addition, an example is provided to show how to directly apply the fifth-order solutions to calculate the velocity distribution and surface elevation when necessary parameters are given.